\documentclass[conference]{IEEEtran}
\IEEEoverridecommandlockouts
\usepackage{cite}
\usepackage{amsmath,amssymb,amsfonts}
\usepackage{algorithmic}
\usepackage{graphicx}
\usepackage{textcomp}
\usepackage{xcolor}
\usepackage{booktabs}

\usepackage[dvipsnames]{xcolor}
\usepackage[final]{microtype}
\usepackage[italic]{mathastext}
\usepackage{libertine}
\usepackage[T1]{fontenc}
\usepackage[varqu,varl]{zi4}
\usepackage[all]{nowidow}
\usepackage{fancyhdr}

\usepackage{booktabs}
\usepackage{array}
\usepackage{tabularx}
\usepackage{longtable}
\usepackage{multirow}
\usepackage{url}
\usepackage[hidelinks]{hyperref}  
\usepackage{enumitem}
\usepackage[most]{tcolorbox}
\usepackage{pifont}
\usepackage{caption}
\def\BibTeX{{\rm B\kern-.05em{\sc i\kern-.025em b}\kern-.08em
    T\kern-.1667em\lower.7ex\hbox{E}\kern-.125emX}}
\begin{document}

\title{DABench-LLM: Standardized and In-Depth Benchmarking of Post-Moore Dataflow AI Accelerators for LLMs\\

\thanks{This work was supported in part by the U.S. National Science Foundation under awards IIS-2416895, IIS-2202699, and OAC-2348465; the U.S. Department of Energy Office of Science, Office of Basic Energy Sciences Data, Artificial Intelligence and Machine Learning at DOE Scientific User Facilities program under Award Number 08735 (``Actionable Information from Sensor to Data Center"), and (DOE) Office of Science, Advanced Scientific Computing Research and Basic Energy Sciences, Advanced Scientific Computing Research for DOE User Facilities award X-ray \& Neutron Scientific Center for Optimization, Prediction, \& Experimentation (XSCOPE).
}
}
\author{
\IEEEauthorblockN{ Ziyu Hu\textsuperscript{*}}
\IEEEauthorblockA{\textit{Dept. of Computer Science}\\
\textit{Stevens Institute of Technology}\\
Hoboken, USA \\
zhu31@stevens.edu} 
\\

\IEEEauthorblockN{Zhijing Ye}
\IEEEauthorblockA{\textit{Dept. of Computer Science}\\
\textit{Stevens Institute of Technology}\\
Hoboken, USA \\
zye25@stevens.edu}

\\
\IEEEauthorblockN{Zheng Xie}
\IEEEauthorblockA{\textit{Dept. of Computer Science}\\
\textit{Binghamton University}\\
Binghamton, USA \\
zxie3@binghamton.edu}

\and
\IEEEauthorblockN{Zhiqing Zhong\textsuperscript{*}}
\IEEEauthorblockA{\textit{Dept. of Computer Science}\\
\textit{Stevens Institute of Technology}\\
Hoboken, USA \\
zzhong9@stevens.edu}
\\
\IEEEauthorblockN{Xuwei Tan}
\IEEEauthorblockA{\textit{Dept. of Computer Science and Engineering}\\
\textit{The Ohio State University}\\
Columbus, USA \\
tan.1206@osu.edu}
\\
\IEEEauthorblockN{Rajkumar Kettimuthu}
\IEEEauthorblockA{\textit{Data Science and Learning Division}\\
\textit{Argonne National Laboratory}\\
Lemont, USA \\
kettimut@anl.gov}
\and

\IEEEauthorblockN{Weijian Zheng}
\IEEEauthorblockA{\textit{Data Science and Learning Division}\\
\textit{Argonne National Laboratory}\\
Lemont, USA \\
wzheng@anl.gov}
\\

\IEEEauthorblockN{Xueru Zhang}
\IEEEauthorblockA{\textit{Dept. of Computer Science and Engineering}\\
\textit{The Ohio State University}\\
Columbus, USA \\
zhang.12807@osu.edu}
\\
\IEEEauthorblockN{Xiaodong Yu}
\IEEEauthorblockA{\textit{Dept. of Computer Science}\\
\textit{Stevens Institute of Technology}\\
Hoboken, USA \\
xyu38@stevens.edu}
}






\maketitle
\begingroup
\renewcommand\thefootnote{*}
\footnotetext{These authors contributed equally to this work.}
\endgroup
\begin{abstract}
The exponential growth of large language models (LLMs) has outpaced the capabilities of traditional CPU and GPU architectures due to the slowdown of Moore’s Law. Dataflow AI accelerators present a promising alternative; however, there remains a lack of in-depth performance analysis and standardized benchmarking methodologies for LLM training. We introduce DABench-LLM, the first benchmarking framework designed for evaluating LLM workloads on dataflow-based accelerators. By combining intra-chip performance profiling and inter-chip scalability analysis, DABench-LLM enables comprehensive evaluation across key metrics such as resource allocation, load balance, and resource efficiency. The framework helps researchers rapidly gain insights into underlying hardware and system behaviors, and provides guidance for performance optimizations. We validate DABench-LLM on three commodity dataflow accelerators, Cerebras WSE-2, SambaNova RDU, and Graphcore IPU. Our framework reveals performance bottlenecks and provides specific optimization strategies, demonstrating its generality and effectiveness across a diverse range of dataflow-based AI hardware platforms.
\end{abstract}


\section{Introduction}
The rapid adoption of decoder-only Large Language Models (LLMs), such as GPT-4~\cite{achiam2023gpt}, LLaMA~\cite{touvron2023llama,grattafiori2024llama,meta2025llama}, and Gemini~\cite{team2024gemini}, has driven exponential growth in computational demands. Meanwhile, the slowdown of Moore’s Law~\cite{lundstrom2022moore} has constrained the performance scaling of traditional architectures, prompting the development of specialized AI accelerators featuring next-generation architectural designs optimized for LLM workloads.

Traditional general-purpose accelerators, such as GPUs, are based on von Neumann architectures that follow an instruction-driven, Bulk Synchronous Parallel (BSP)-style execution model~\cite{backus1978can}. When running LLM workloads on such architectures, frequent data access and the transfer of large model parameters often lead to severe memory bottlenecks, low compute utilization, and poor energy efficiency~\cite{gholami2024ai}. In contrast, many emerging AI accelerators adopt dataflow architectures that employ a data-driven execution model, in which processing elements are asynchronously triggered as soon as all required input data becomes available, eliminating the need for global instruction scheduling. This execution paradigm aligns naturally with the structure of LLM workloads and can substantially improve compute resource utilization and parallelism by avoiding the global synchronization and excessive memory access overhead inherent in conventional architectures~\cite{sambanova2021whitepaper}.

Although dataflow architectures hold great promise for LLM workloads, their practicality and performance characteristics remain underexplored. Existing benchmarking studies on emerging AI accelerators tend to focus either on chip design metrics, such as chip area and memory capacity~\cite{kundu2025comparison}, or on high-level performance metrics, such as the throughput of neural network building blocks (e.g., convolutional kernels), across various hardware platforms~\cite{emani2022comprehensive, peng2024evaluating, shekofteh2023performance}. A few studies~\cite{emani2024toward, john2024performance, zhang2024benchmarking} examine LLM workloads (e.g., BERT or GPT) on dataflow accelerators, but they primarily investigate the impact of model configurations on performance and report only high-level metrics (e.g., TFLOPs). While these works offer useful insights into the overall performance of AI accelerators, they lack a systematic, full-stack analysis that connects system-level characteristics to the execution performance of LLM workloads. Compounding this challenge, commodity dataflow AI accelerators often incorporate diverse vendor-specific designs and optimizations, the details of which are rarely made public due to business constraints. This lack of a generic and in-depth understanding significantly limits the research community's ability to comprehensively evaluate and optimally utilize such accelerators, particularly in the context of LLM workloads.

To fill this gap, we propose DABench-LLM, a unified framework for performing standardized and in-depth benchmarking of both existing and future dataflow AI accelerators, with minimal vendor-specific adaptations. The framework provides a generalizable guideline to help researchers rapidly gain insights into key hardware and system characteristics, and systematically uncover the execution strategies, scheduling mechanisms, and performance bottlenecks of dataflow-based AI accelerators. Specifically, the framework operates on two levels: (1) At the chip level, it analyzes how dataflow accelerators execute LLM workloads by examining resource allocation rates, load balancing, and the utilization of memory and compute resources. (2) At the multi-chip level, it evaluates the scalability of these accelerators and investigates deployment strategies that affect overall system performance and optimization opportunities. The first level offers complete insights to guide kernel and graph compiler optimizations for dataflow AI chips, while the second level informs communication and model deployment strategies on platforms equipped with dataflow-based AI accelerators.

To validate and showcase the effectiveness of DABench-LLM, we evaluate it on three commercially available dataflow accelerators - Cerebras, SambaNova, and Graphcore - provided through the AI testbed\footnote{https://www.alcf.anl.gov/alcf-ai-testbed} at the Argonne Leadership Computing Facility (ALCF), and conduct additional experiments on the Neocortex system at the Pittsburgh Supercomputing Center (PSC). Although all three adopt dataflow architectures, they differ significantly in their vendor-specific chip designs and execution models. Specifically, Cerebras aims to accommodate the entire model on a single, extremely large chip. SambaNova partitions the computation graph into multiple sections, executing them sequentially on a single chip. Graphcore employs pipeline parallelism, grouping layers and distributing them across multiple chips to form a distributed execution pipeline. Despite these differences, our framework is readily applicable to all three accelerators. It successfully delivers benchmarking results and insights that can adequately guide further optimizations. These case studies highlight the generality and effectiveness of our framework in evaluating a wide range of dataflow-based AI hardware platforms. Our code is publicly available at \url{https://github.com/augustuszzq/Regular-DABench-LLM}.

Our contributions in this work can be summarized as:
\begin{enumerate}
  \item We propose DABench-LLM, the first framework that enables standardized and in-depth benchmarking of diverse dataflow AI accelerators running LLMs. It helps researchers rapidly gain hardware and system insights and provides guidance for performance optimizations.
  \item We validate and showcase DABench-LLM's effectiveness on three commodity dataflow AI accelerators, providing comprehensive benchmarking results that reveal their hardware and system characteristics. 
  \item We conduct in-depth analyses based on DABench-LLM results to extract key insights and offer optimization guidelines for each accelerator.
\end{enumerate}

The remainder of the paper is organized as follows: Sec. 2 provides background, Sec. 3 discusses our insights into mapping LLM workloads onto various AI chips, Sec. 4 elaborates on the design of the DABench-LLM framework, and Sec. 5 and 6 demonstrate the effectiveness of DABench-LLM on three commodity dataflow AI accelerators. Sec. 7 and 8 cover related work and conclude the paper, respectively.
\section{Background}

\subsection{Decoder-Based Large Language Models}
Most state-of-the-art large language models (LLMs), such as Llama\cite{meta2025llama} and GPT\cite{achiam2023gpt}, adopt a decoder-only architecture. Compared to encoder-based models like T5 and BERT, decoder-only models typically achieve better generalization\cite{wang2022language}. This architecture offers several advantages: prompts directly influence all decoder layers, enhancing fine-tuning signals and enabling superior in-context learning\cite{dai2022can}; additionally, causal attention provides implicit positional encoding, addressing the position-invariance issue of standard Transformers\cite{haviv2022transformer}. These benefits have led to the widespread adoption of decoder-only designs in large-scale language modeling.

GPT-2\cite{radford2019language} is an early, influential decoder-only Transformer language model that uses learnable absolute positional embeddings, GELU activation functions, and multi-head self-attention with residual connections and layer normalization; typical configurations span 117M–1.5B parameters with approximately 1024-token context length, and it remains widely used as a baseline model for comparative studies and educational instruction due to its relatively modest computational requirements and well-understood behavior. LLaMA-2\cite{touvron2023llama} represents a modern, open-source, commercially usable decoder-only family (7B/13B/70B parameters) that employs RoPE (Rotary Position Embedding) for improved length extrapolation, RMSNorm for enhanced training stability, and SwiGLU activation functions for better expressivity, with grouped-query attention implemented in larger variants to reduce memory bandwidth requirements; it typically supports approximately 4096-token context length and offers both base models for general-purpose fine-tuning and dialogue-optimized variants (-Chat, trained with supervised fine-tuning and reinforcement learning from human feedback alignment).Given that GPT-2 and LLaMA-2 are canonical decoder-only architectures with comprehensive community support and broad compatibility across diverse AI accelerators, we use these two representative model families as the primary basis for our systematic experiments in this paper.
\subsection{Specs of Commodity Dataflow AI Hardware}
To demonstrate the applicability of our benchmarking framework, we evaluate three dataflow-based AI hardware platforms—Cerebras, SambaNova, and Graphcore—and measure their performance using our framework. In this subsection, we briefly introduce their specifications based on the vendors’ white papers.
\subsubsection{Cerebras}
\begin{figure}[htbp]
    \centering
    \includegraphics[width=0.75\linewidth]{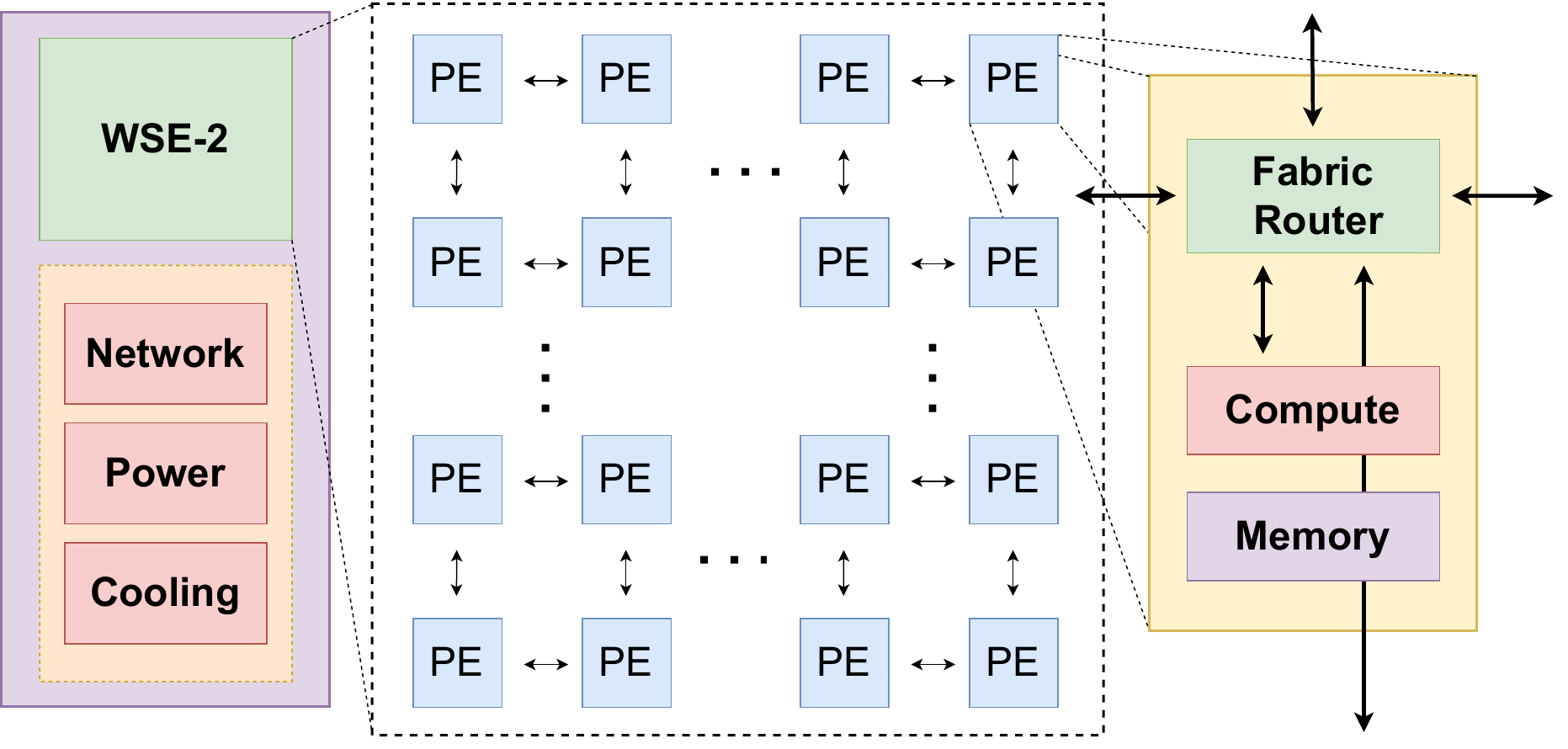}
    \caption{Architecture of CS-2 System. The left side shows the overall architecture of the CS-2 system, the middle illustrates the structure of its core component—the WSE-2 chip, and the right side presents the architecture of the basic computational unit (PE, Processing Element) that composes the WSE-2.}
    \label{fig:example_cs2}
\end{figure}
In this work, we utilize the CS-2 system—Cerebras’ previous-generation AI accelerator—which integrates cooling infrastructure, network and power interfaces, and its core processor, the WSE-2 (Wafer-Scale Engine 2) (Figure \ref{fig:example_cs2}). While the newer CS-3 system adds external memory modules to the WSE-2 architecture, we focus on CS-2 due to the lack of publicly available chip-level data for CS-3. The WSE-2 processor features 850,000 processing elements (PEs), each equipped with a dedicated Sparse Linear Algebra Compute (SLAC) core optimized for executing the sparse linear algebra operations fundamental to neural network workloads. The chip includes 40GB of on-chip memory uniformly distributed across all PEs, with each PE capable of accessing up to 48KB of local SRAM. This architecture enables a total on-chip memory bandwidth of 20PB/s. Inter-PE communication is supported by the Swarm communication fabric, which connects each PE to its five neighboring directions: the four adjacent PEs in the cardinal directions (north, south, east, and west), as well as its local compute and memory unit, yielding a total communication bandwidth of 220PB/s \cite{cerebras2021datasheet}.
\subsubsection{SambaNova}
\begin{figure}[htbp]
    \centering
    \includegraphics[width=0.85\linewidth]{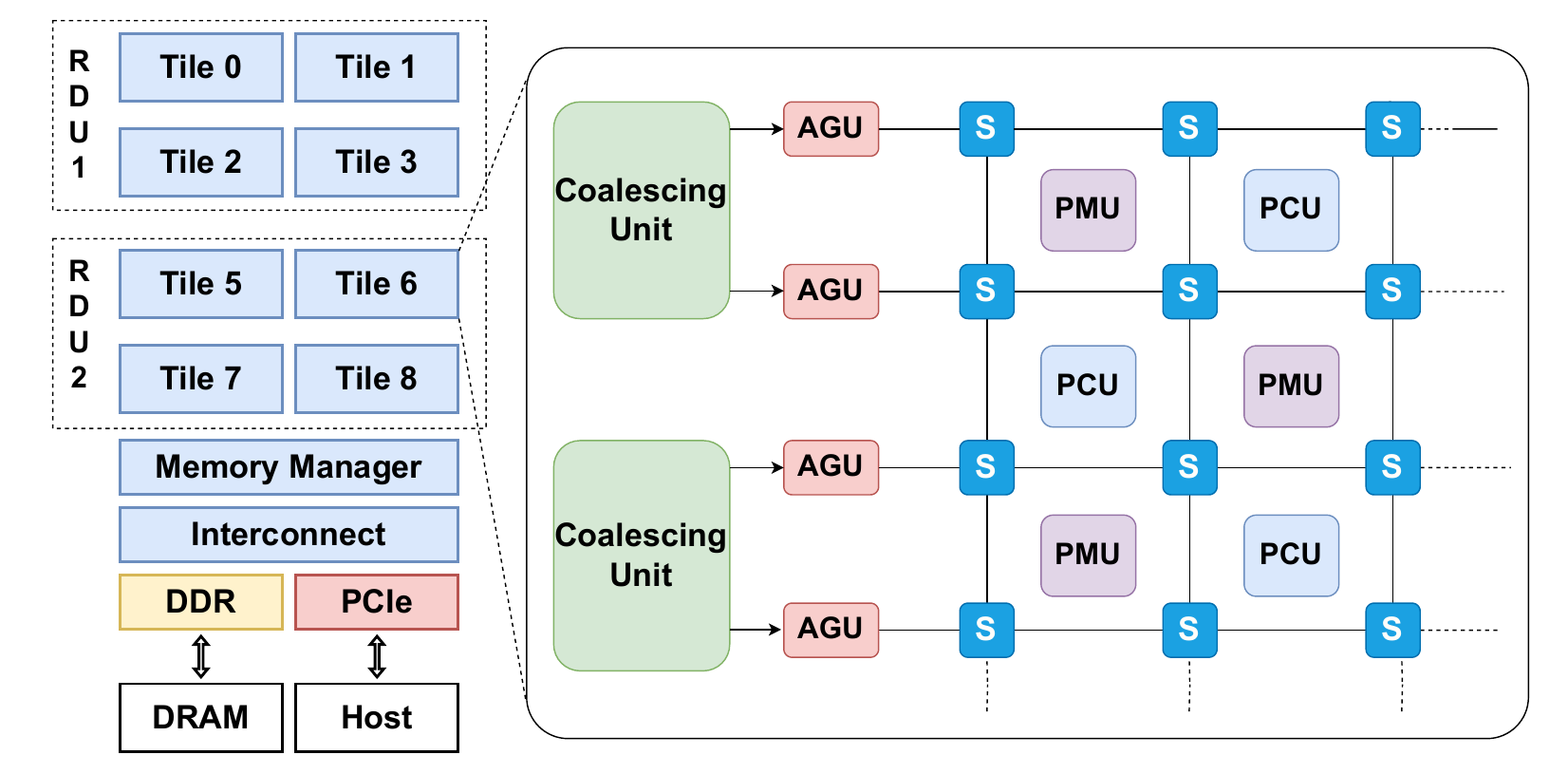}
    \caption{Architecture of SN30 system. 
    The left side shows two compute units (RDU1 and RDU2) with tiles connected to memory and host through a memory manager and interconnect.
    The right side is a zoom-in view of an RDU, showing the internal grid of switches connecting processing units and address generators.}
    \label{fig:example_rdu}
\end{figure}
We utilize the SambaNova DataScale SN30 system, which contains two Reconfigurable Dataflow Units (RDUs) (Figure \ref{fig:example_rdu}). Each RDU consists of four tiles built on a modular dataflow architecture, interconnected via a unified memory controller and a top-level communication fabric. RDUs interface with off-chip DDR memory and the host processor through PCIe. Within each tile, 160 Pattern Compute Units (PCUs) enable massively parallel SIMD execution, while 160 Pattern Memory Units (PMUs) function as intelligent scratchpads to reduce off-chip memory traffic. On-chip communication is facilitated by Switch (S) modules, which coordinate a 3D switching fabric composed of scalar, vector, and control networks to support both inter-tile and inter-RDU data transfer. Additionally, Address Generator Units (AGUs) and Coalescing Units (CUs) enable high-bandwidth access to DRAM, other RDUs, and the host via the RDU Connect interface, with native support for sparse and graph-structured data \cite{sambanova2021whitepaper}.

\subsubsection{Graphcore}
\begin{figure}[htbp]
    \centering
    \includegraphics[width=0.75\linewidth]{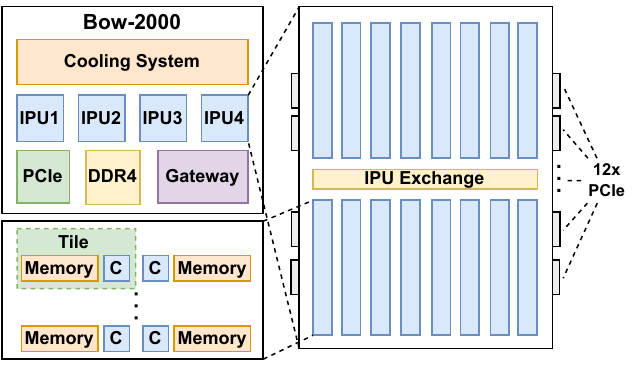}
    \caption{Architecture of Bow-2000. The top-left panel illustrates the overall architecture of the Bow-2000 system. The right panel presents the internal structure of an IPU, while the bottom-left panel shows the layout of a tile column and the basic structure of a single tile.}
    \label{fig:ipu_hard}
\end{figure}

Each Bow-2000 system, as shown in the top-left of the Figure \ref{fig:ipu_hard}, integrates a cooling system, a PCIe interface, and four IPUs connected via a Gateway \cite{graphcore2025bow2000}. These IPUs share 256GB of external DDR memory. As illustrated on the right, each IPU comprises 1,472 tiles in total, along with 12 PCIe interfaces and an all-to-all IPU-Exchange interconnect offering 8TB/s bandwidth. Each tile contains an IPU core (C) and local memory, as illustrated in the bottom-left of the figure. The IPU cores follow a multi-instruction, multi-data (MIMD) model, supporting up to six concurrent threads per core with 64KB of shared on-tile memory. Tile-to-tile communication is enabled through the IPU-Exchange, while host communication is handled via the PCIe links \cite{graphcore2025ipu}.
\section{Insights for Mapping LLMs onto Chips}

To investigate the execution of large language model (LLM) training across diverse hardware platforms, we base our analysis primarily on empirical observations, complemented by insights from prior research and official documentation.
In dataflow systems, programs are typically represented as computation graphs, where nodes denote operators and edges represent data dependencies. These graphs are then mapped onto the chip and executed asynchronously. 

The following sections describe the mapping and execution strategies adopted by each type of accelerator.
\subsection{Cerebras}
In Cerebras CS-2, the entire LLM model is represented as a computation graph constructed at the granularity of individual layers and is loaded onto the WSE-2 chip as a whole. Each layer in the model is mapped to a kernel, and each kernel is assigned a certain number of processing elements (PEs) to execute its computation. The performance of a kernel largely depends on the number of PEs allocated to it. For example, in a multi-decoder model, each attention layer is compiled into a separate attention kernel; all attention kernels are deployed onto the WSE-2 chip and are assigned an equal or similar number of PEs. Typically, kernels with data dependencies are placed physically close to each other on the chip to reduce communication overhead. Once data is injected from the embedding layer, it propagates through the computation graph in a data-driven manner, with each subsequent computation triggered by data availability, eliminating the need for global instruction scheduling.

\subsection{SambaNova}
On the SN30 system, the entire LLM workload is first represented as a computation graph, where nodes correspond to operators and edges denote data dependencies. This graph is then partitioned into multiple subgraphs, referred to as sections, which are sequentially loaded onto a RDU for execution. Each section is processed one at a time. All model parameters and intermediate data are stored in off-chip DDR memory, and during execution, the required data is transferred to the on-chip Pattern Memory Units (PMUs). Depending on the graph partitioning strategy, SN30 supports three compilation modes: O0 (Operator mode), O1 (Module mode), and O3 (Full graph mode).
\begin{figure}[htbp]
    \centering
    \includegraphics[width=0.7\linewidth]{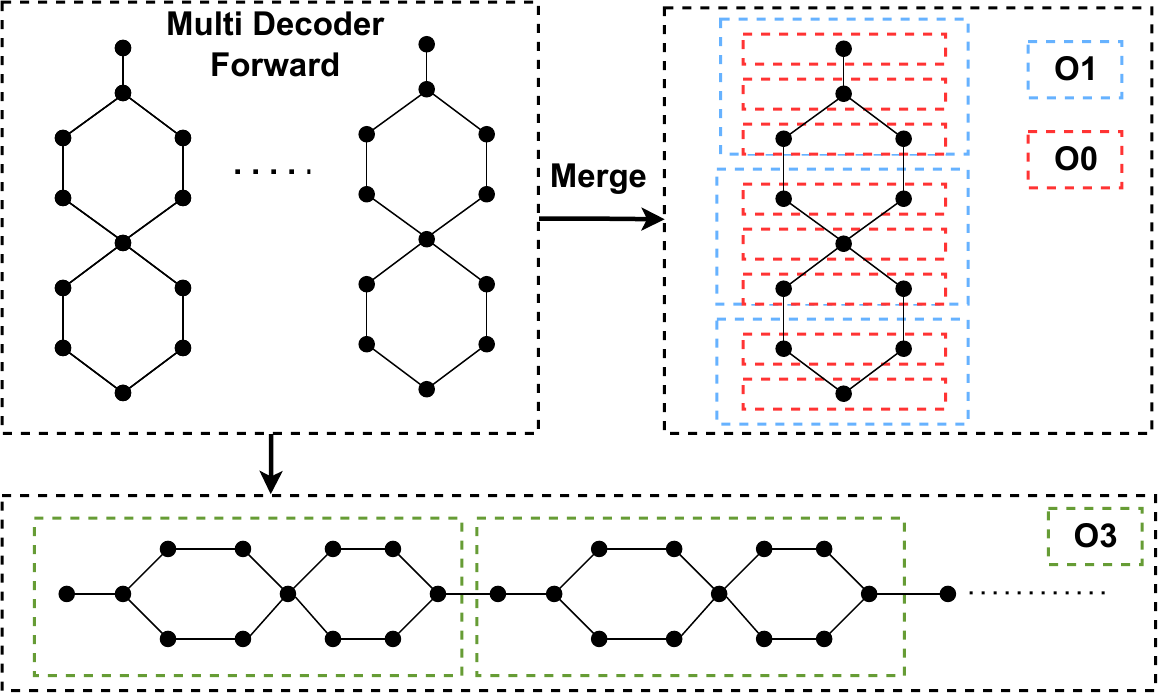}
    \caption{Section Partitioning Strategies under Different Compilation Modes (Example: Decoder Forward Pass)}
    \label{fig:compile}
\end{figure}

Using the forward pass of all decoder layers in an LLM as an example, Figure \ref{fig:compile} illustrates how sections are partitioned under different compilation modes. Red boxes represent sections in O0 mode, blue boxes correspond to O1 mode, and green boxes indicate O3 mode. The top-left of the figure shows the original computation graph of multiple decoder layers. In both O0 and O1 modes, all decoder layers are merged into a single unified subgraph (shown in the top-right). In O0 mode, each operator is individually assigned to a section; in O1 mode, SambaNova's operator fusion strategy is applied to group multiple operators into larger modules, which are then partitioned into sections. During execution, each section is invoked multiple times depending on the number of decoder layers, and the forward computations across all layers proceed in parallel and synchronously. In contrast, O3 mode retains the independence of each decoder layer and partitions them sequentially into separate sections without fusion. In this mode, section boundaries do not strictly align with individual decoders but are adjusted dynamically based on model size. Execution proceeds in a decoder-by-decoder manner across layers.

\subsection{Graphcore}

Graphcore’s architecture is inherently designed to support pipeline parallelism (PP). Its execution model partitions the computation graph by layers and maps each part onto different IPUs for execution. Typically, the embedding layer is assigned to a dedicated IPU, while the decoder layers are grouped based on the number of remaining IPUs and distributed across them. The IPUs are organized into a pipeline according to the computational dependencies. Training a language model therefore requires at least two IPUs—one for the embedding layer and another for the decoder layers.

\section{Framework Design Methodology}

\subsection{Framework Overview}
\begin{figure}[htbp]
    \centering
    \includegraphics[width=0.92\linewidth]{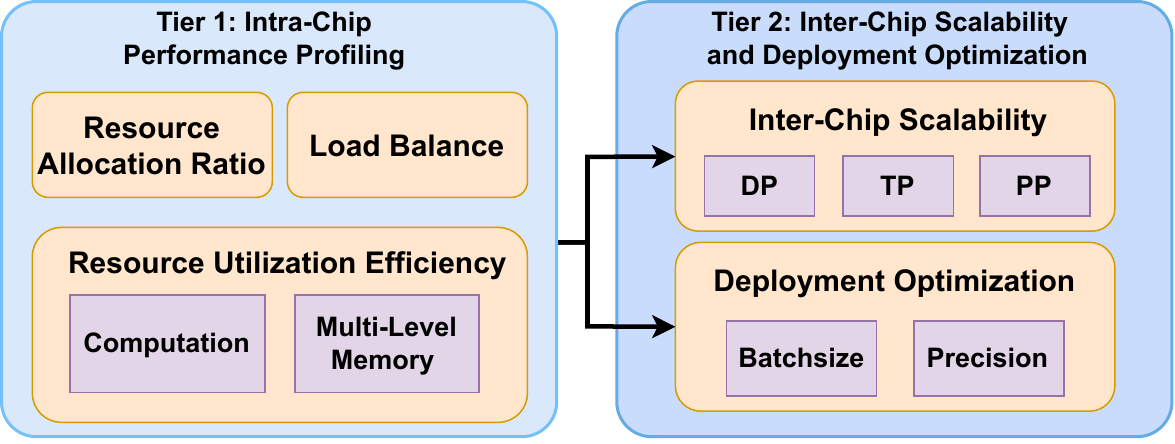}
    \caption{Overview of framework. Given a new dataflow-based hardware platform, Tier-1 evaluates its performance when executing LLM workloads on a single chip, while Tier-2 assesses its scalability across multiple chips and investigates optimization strategies for deployment. }
    \label{fig:framework}
\end{figure}

As most AI accelerator ecosystems today remain relatively closed, the performance and effectiveness of various hardware platforms often depend on vendor-provided benchmarks. To gain deeper insights into how these dataflow-based AI chips perform in large language model (LLM) training, we develop a unified evaluation framework, as illustrated in Figure \ref{fig:framework}. The framework employs a systematic two-tier approach:
\begin{itemize}
\item \textbf{Tier 1: Intra-Chip Performance Profiling} 
- This tier characterizes the theoretical performance upper bound of the chip through resource allocation ratios, assesses resource waste via load balance analysis, and evaluates resource utilization efficiency to measure how effectively the chip leverages compute units and multilevel memory resources (e.g., on-chip and off-chip memory) under actual LLM workloads.
\item \textbf{Tier 2: Inter-Chip Scalability and Deployment Optimization} - This tier evaluates the scalability of the accelerator under real-world LLM workloads—where growing model sizes make scalability essential—and examines deployment optimization strategies such as batch size selection, precision modes, and other performance-critical factors.
\end{itemize}

To enable comprehensive evaluation across these two tiers, our benchmark design adopts a decoder-block approach as the fundamental evaluation unit, recognizing that full-scale LLMs are impractical for single-chip analysis on current dataflow accelerators. By systematically varying layer count and hidden size parameters, this methodology probes different performance bottlenecks across the compute-memory spectrum, ensuring comprehensive coverage of diverse computational intensity and memory access patterns. 

\subsection{Tier1: Intra-Chip Benchmarking}

In characterizing each chip, our framework focuses on three key metrics: resource allocation ratio, load balance, and resource utilization efficiency.

\textit{Resource allocation ratio} refers to the proportion of resource involved in the LLM workload relative to the total available resource on the chip. This ratio represents the upper bound of the chip’s achievable efficiency.

\textit{Load balance} describes the degree of uniformity in resource allocation during the execution of LLM workloads. In dataflow architectures, an LLM task is divided into multiple subtasks that are mapped to different regions of the chip, with each subtask triggered by the output of its predecessor. As a result, overall throughput is typically limited by the slowest subtask. The granularity of task partitioning affects this metric—finer granularity provides a more accurate indication of resource inefficiency.

\textit{Resource utilization efficiency} refers to the actual usage of compute resources and multilevel memory resources by the chip during the execution of LLM workloads, reflecting the overall performance of the chip under real-world conditions.

\subsubsection{Resource Allocation Ratio}
The resource allocation ratio measures how many on-chip units the compiler allocates to a workload, which is defined as: 
\begin{equation}
\text{U} =  \frac{R_{used}}{R_{all}} 
\end{equation}
where $R_{used}$ is the number of units used by the workload, and $R_{all}$ is the total number of available units. 

In WSE-2 and IPU architectures, compute and memory resources are respectively associated with PEs and tiles. In the case of RDU, where the training of a language model is divided into multiple sections, we use the execution time of each section as a weight to compute the weighted average of PCU and PMU utilizations, representing the overall resource allocation ratio. The computation is defined as follows:
\begin{equation}
\text{U} = \frac{\sum_{i=1}^{N} L_i \cdot \left( \frac{R_i}{R_{all}} \right)}{\sum_{i=1}^{N} L_i}
\end{equation}
where $L_i$ is the runtime of the $i$-th section, $R_i$ is the number of units it uses (either PCUs or PMUs).

\subsubsection{Load Balance}
Since the slowest task determines the overall system throughput, other tasks inherently experience some degree of resource underutilization—the greater the throughput gap, the more severe the inefficiency. To quantify this phenomenon, we introduce the Load Imbalance (LI) metric, which measures the degree of throughput disparity across tasks, defined as follows:

\begin{equation}
\text{LI} = \frac{1}{\sum_{i=1}^{N} R_i} \sum_{i=1}^{N} \left(\frac{T_{\min}}{T_i} \cdot R_i\right)
\end{equation}

where $N$ represents the number of tasks, $R_i$ denotes resources allocated to task $i$, $T_i$ is the throughput of task $i$, and $T_{\min}$ denotes the minimum throughput across all tasks. An LI value closer to 0 indicates higher imbalance, while a value closer to 1 indicates better balance.

For the RDU, we likewise compute the overall LI using a time-weighted average across sections:

\begin{equation}
\text{LI}_{\text{total}} = \frac{\sum_{i=1}^{N} L_i \cdot \text{LI}_i}{\sum_{i=1}^{N} L_i}
\end{equation}

where $N$ is the total number of sections, $L_i$ is the runtime of section $i$, and $\text{LI}_i$ is the load imbalance value of section $i$.

\subsubsection{Resource Utilization Efficiency}

This metric includes the chip’s compute performance during LLM execution (TFLOPs) and the utilization and bandwidth of its multi-level memory. As all three accelerators have at most two memory levels, we adopt the GPU-style classification and refer to them as \textit{shared memory} and \textit{global memory} tiers in this paper.

The \textit{Roofline Model} is used as an auxiliary tool in this investigation to visualize the interaction between arithmetic intensity and achievable throughput. Due to the lack of publicly available bandwidth data for shared memory, the model is applied only to the global memory tier.
\paragraph{Arithmetic Intensity Estimation for LLM Training}
We estimate the arithmetic intensity (AI) of transformer-based LLM training workloads using the following expression:
\begin{equation}
AI = \frac{6 \times P \times B \times S}{4 \times P + \text{Activation Memory}}
\end{equation}
where $P$ denotes the number of model parameters, $B$ is the batch size, and $S$ is the input sequence length. The constant factor 6 accounts for both forward ($2 \times$) and backward ($4 \times$) FLOPs per token. The denominator estimates total memory traffic, including both model weights and intermediate activations.~\cite{wang2019benchmarking, jia2019dissecting, narayanan2021efficient}

\subsection{Tier 2: Inter-Chip Benchmarking}
\subsubsection{Inter-Chip Scalability}
The inter-chip scalability of dataflow-based accelerators should be analyzed through the lens of the three classical parallelism strategies used in GPUs: \textit{data parallelism (DP)}, \textit{pipeline parallelism (PP)}, and \textit{tensor parallelism (TP)}. Although emerging hardware platforms rarely adopt these strategies in a direct manner, their scalability mechanisms are often inspired by them and shaped by their unique architectural constraints. Our framework adopts this classification as a conceptual basis to characterize the scalability patterns of different accelerators and uses it to analyze the resulting communication and other associated overheads.
\subsubsection{Deployment Optimization}
Maximizing throughput is the primary motivation for employing accelerators in LLM workloads. Among the many factors that affect throughput, \textit{batch size} and \textit{precision} exert the most significant impact across most modern accelerators. As their effects vary by hardware platform, these two factors should be prioritized in deployment decisions to achieve throughput-optimized execution.

\subsection{Framework Validation Designs}

We validate our framework on three representative dataflow accelerators: CS-2, SN30, and Bow-2000. Our goals are to uncover platform-specific bottlenecks, demonstrate cross-architecture generalizability and offer deployment insights. Importantly, we do not compare different hardware platforms because ensuring fairness is highly challenging.

\paragraph{Methodology and Setup}
Full-scale LLMs are impractical on a single chip, so we adopt a decoder-block approach. By fixing hidden size (HS) or layer count, we probe compute, memory, and communication limits. GPT-2 and LLaMA2 experiments yield key framework metrics. In the intra-chip section, the GPT-2 Small decoder block (hidden size 768) is our basic unit on Cerebras and Graphcore; on SambaNova we use GPT-2 for O0/O3 and the LLaMA-2 decoder block for O1. In the inter-chip section, we used GPT mini, tiny, and small models on WSE-2 (hidden sizes 256, 512, and 768), the LLaMA-2 7B model on the RDU, and a GPT model with a hidden size of 768 on the IPU.
\paragraph{Required Information}
To run our framework, the information to be collected mainly falls into three categories: hardware specifications, including the number of on-chip units, total memory capacity, and bandwidth at each memory level, etc.; runtime information, including per-task throughput, the number of units allocated to each task, achieved TFLOPs, memory usage, and overall throughput, etc.; and training configuration, including batch size, sequence length, model parameter count, degree of parallelism, numerical precision, etc.

\paragraph{Platform Constraints}Platform constraints make it difficult to collect runtime data in a unified manner. For WSE, RDU-O1, and IPU, most metrics are from compile time, with only a few (e.g., throughput, TFLOPs) from runtime. In contrast, RDU-O0 and O3 rely entirely on runtime data. We prioritize runtime data and apply weighting to reduce batch variance on sensitive systems like CS-2, ensuring fair cross-platform comparisons. Since most metrics are determined at compiling time and remain unchanged during execution, the framework reliably captures actual performance characteristics.
\begin{figure*}[htbp]
    \centering
    \begin{minipage}[t]{0.36\linewidth}
        \centering
        \includegraphics[width=\linewidth,height=0.18\textheight]{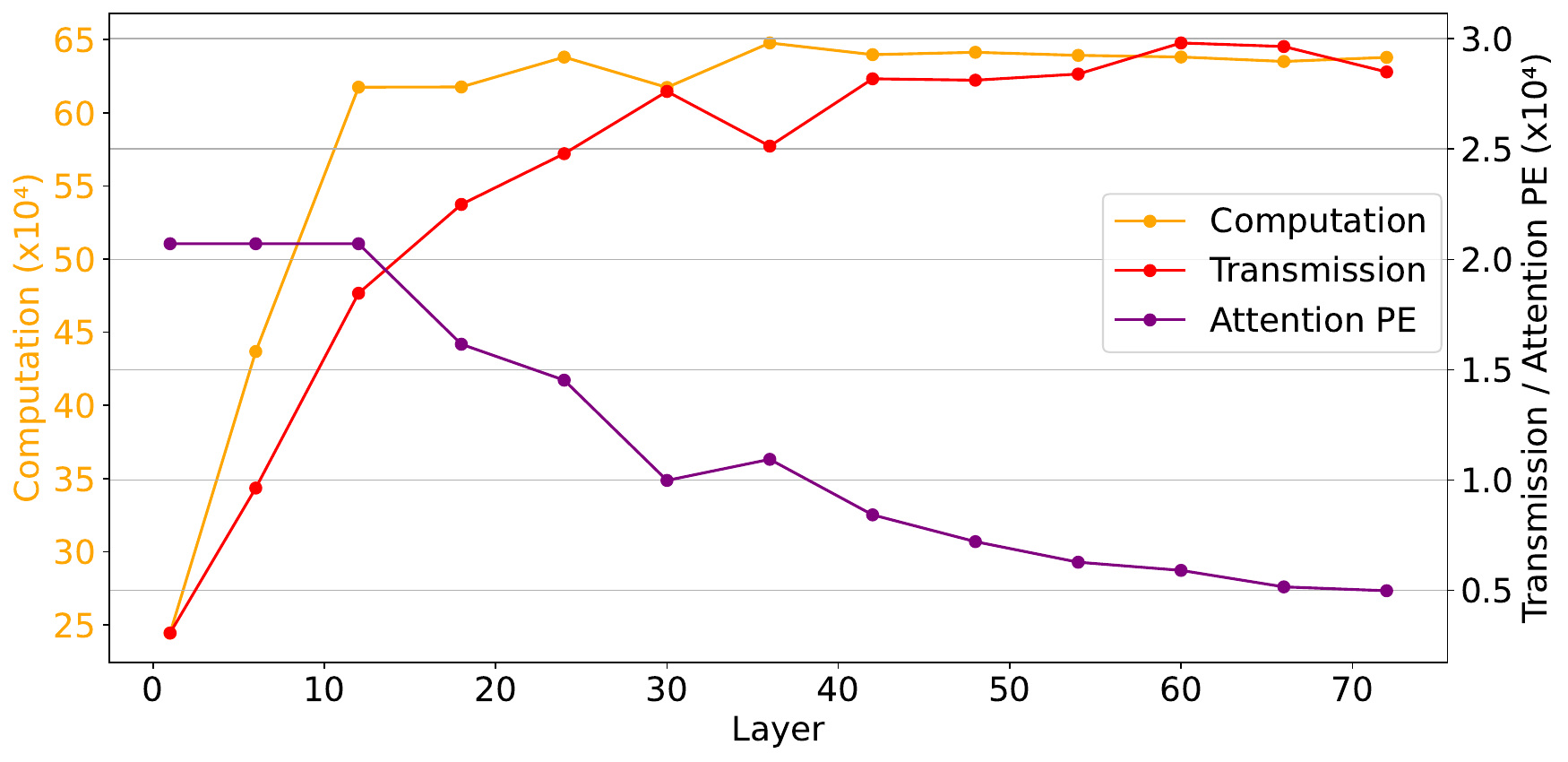}
        \caption{PE Usage Breakdown. As layer count increases, total PEs for computation (orange, left y-axis) and Transmission (red, right y-axis) rise, while PEs per attention kernel (purple, right y-axis) gradually decrease—demonstrating WSE-2's elastic resource scheduling.}
        \label{fig:attention_pe}
    \end{minipage}%
    \hfill
    \begin{minipage}[t]{0.62\linewidth}  
        \centering
        \begin{minipage}[t]{0.485\linewidth}
            \centering
            \includegraphics[width=\linewidth]{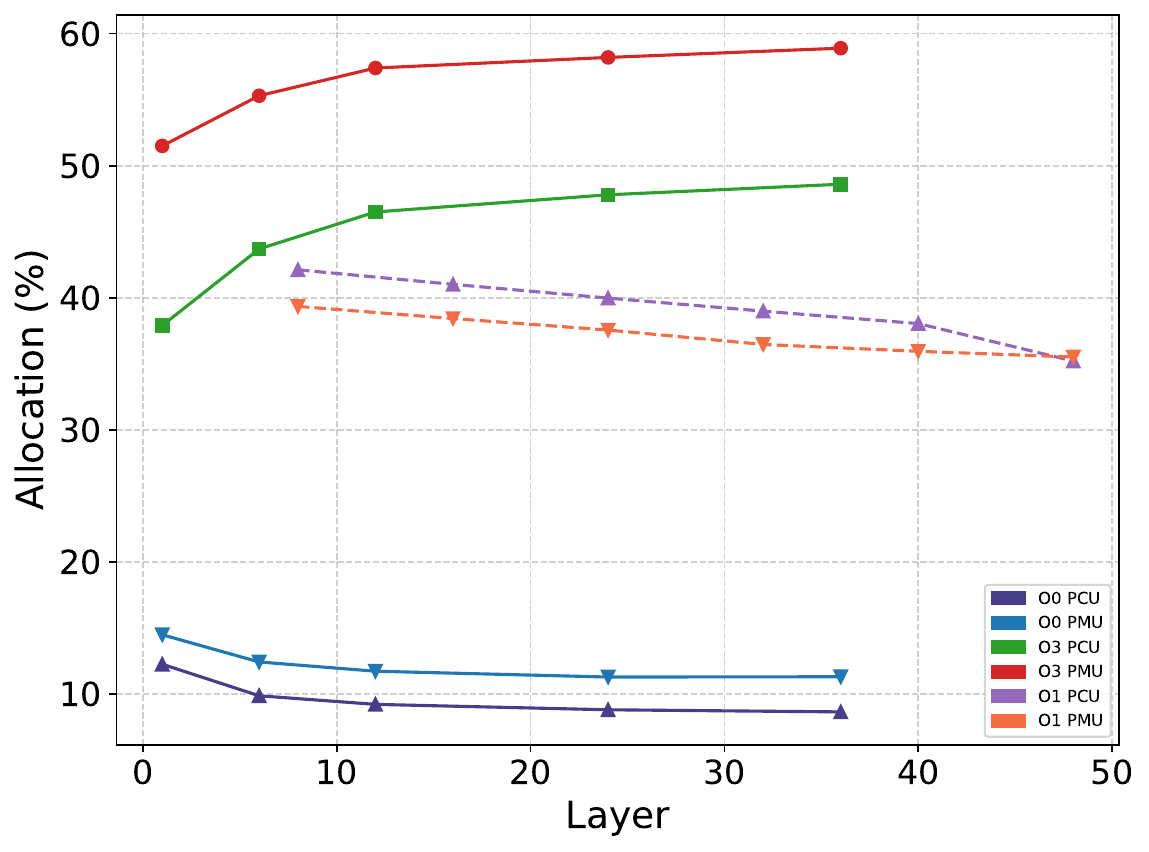}
            \par
            \vspace{1pt}
            \centering
            \footnotesize (a) Allocation vs Layer
        \end{minipage}%
        \hfill
        \begin{minipage}[t]{0.485\linewidth}
            \centering
            \includegraphics[width=\linewidth]{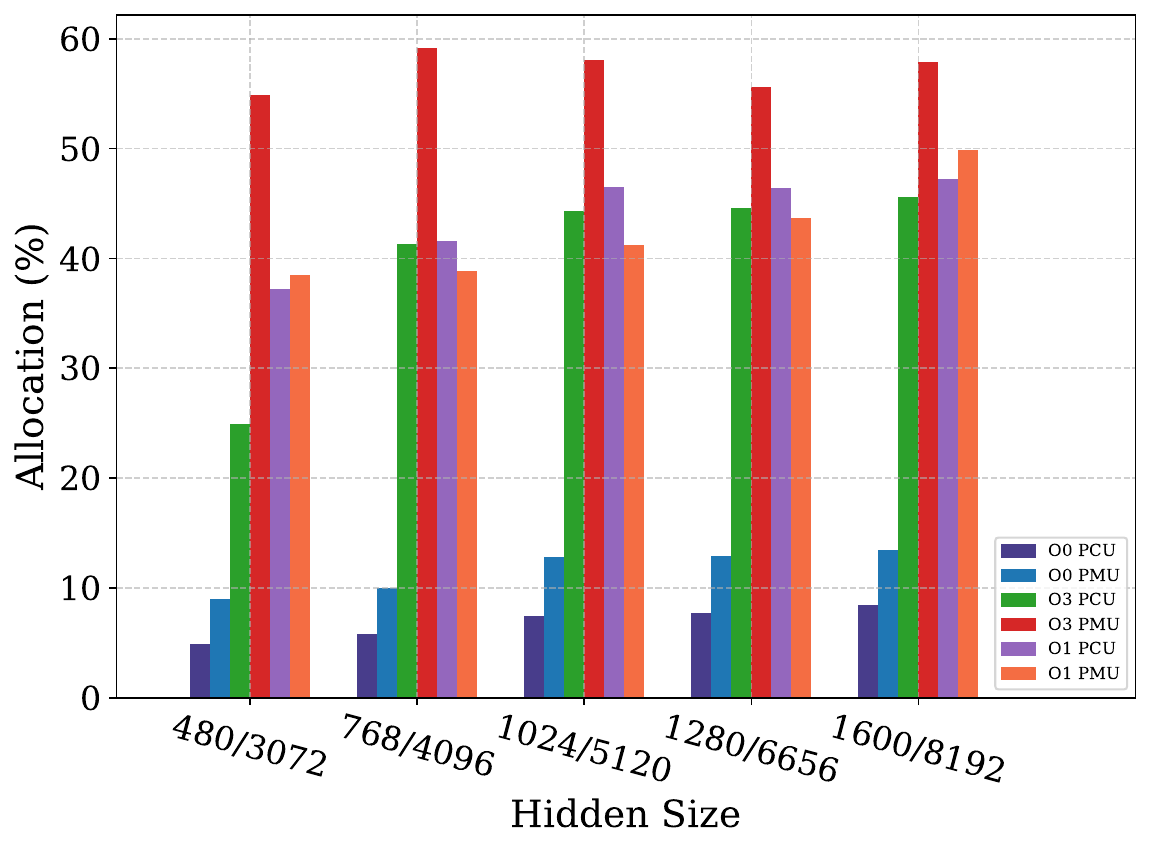}
            \par
            \vspace{1pt}
            \centering
            \footnotesize (b) Allocation vs Hidden Size
        \end{minipage}
        \vspace{3pt}
        \caption{RDU Resource Allocation Ratio Across Layers and HS. Figure (a) shows how the allocation ratio changes with the number of layers; Figure (b) illustrates the impact of HS on resource allocation. O0 and O3 modes use smaller HS (480–1600, left of the x-axis), while O1 mode operates on larger HS (3072–8192, right).}
        \label{fig:pmu_uti}
    \end{minipage}
\end{figure*}

\section{Intra-Chip Performance Profiling}
\label{sec:performance}

This section validates Tier 1 of our framework across the three hardware platforms. We evaluate each chip's performance for different LLM workloads and analyze the underlying bottlenecks.

\subsection{Resource Allocation Analysis}
In dataflow accelerators, resource allocation often defines the upper bound on achievable utilization. WSE-2 and IPU use tightly coupled compute–memory units (PEs and tiles), while RDU separates them into PCUs and PMUs. Due to lack of tile-level data, our analysis focuses on WSE-2 and RDU.
\begin{table}[h!]
\centering
\caption{PE allocation ratio across different layer configurations}
\label{tab:pe_usage}
\renewcommand{\arraystretch}{1.2}
\resizebox{1\linewidth}{!}{%

\setlength{\tabcolsep}{2pt}
\begin{tabular}{c|cccccccccccccc}
\toprule
\textbf{Layer}    & 1 & 6 & 12 & 18 & 24 & 30 & 36 & 42 & 48 & 54 & 60 & 66 & 72 & 78 \\
\midrule
\textbf{Pe(\%)}  & 33 & 60 & 85 & 87 & 91 & 88 & 92 & 92 & 92 & 92 & 92 & 92 & 93 & Fail\\
\bottomrule
\end{tabular}
}
\end{table}

\subsubsection{Cerebras}
Since WSE-2 chip loads the entire computation graph onto the chip without any partitioning, we aim to understand how the PE allocation rate changes with increasing model parameters. To ensure a linear growth in parameter size, we selected the decoder-only model with a HS of 768 and varied the number of layers for our study. This HS was chosen because larger HS make it difficult to observe changes in PE allocation, while smaller dimensions result in overly trivial experiments.

Table~\ref{tab:pe_usage} shows PE allocation increases with layer count up to 36 layers, then stabilizes around 92-93\%, indicating that even under ideal conditions, PE allocation ratio cannot exceed approximately 93\%. 

In WSE-2, PEs are allocated to different tasks, primarily computation and data transmission. We report the number of PEs assigned to computation and to transmission in Figure~\ref{fig:attention_pe}, which reveals detailed PE-allocation patterns: Computation and Transmission PEs follow similar trends with close proportions. For individual attention layers, PE usage decreases as model parameter size increases, demonstrating elastic adaptation to on-chip resource availability. This mechanism enables high PE utilization without manual resource scheduling when computational workload is sufficiently large.

However, each kernel function has an optimal PE allocation threshold. For layers below 12, PE usage per attention kernel remains stable because additional resources beyond this threshold lead to diminishing returns due to increased communication overhead between PEs, as shown in Figure~\ref{fig:attention_pe}. This kernel scalability limitation results in resource underutilization, which explains the relatively low PE allocation observed for layers below 12, as detailed in Table~\ref{tab:pe_usage}.

\begin{tcolorbox}[colback=gray!20, 
  colframe=black, 
  boxsep=2pt,        
  left=2pt,          
  right=2pt,         
  top=2pt,           
  bottom=2pt         
  ]
\textbf{Insight:} WSE-2 achieves a high on-chip resource allocation ratio (92–93\%, Table~\ref{tab:pe_usage}) through its flexible kernel allocation strategy. However, it faces scalability limitations with large models, supporting up to 72 decoder layers ($\sim$500M parameters) in our experiments.
\end{tcolorbox}

\subsubsection{SambaNova}
Due to the complexity and diversity of execution modes of RDU chip from SambaNova, we make our best effort to unify experimental settings. Although some inconsistencies remain, the observed performance trends on the RDU provide more meaningful insights into its scheduling strategies and bottlenecks than fixed numerical comparisons.

Figure~\ref{fig:pmu_uti} shows overall RDU resource allocation never exceeds 60\%, with O3 achieving the highest and O0 the lowest utilization, indicating considerable underutilization when processing LLM tasks. O0 and O1 behave almost identically, while O3 exhibits distinct patterns due to different section partitioning strategies.

O3 mode performs sequential decoder-by-decoder computation, so average allocation approaches single decoder allocation as layers increase. The O3 curve gradually increases and stabilizes, suggesting non-decoder sections have lower resource allocation ratio. In contrast, O0 and O1 execute all decoders synchronously by operator type—when layers increase, higher-intensity sections overlap better in execution, reducing their latency share and lowering average allocation rates.
\begin{table}[h!]
\caption{O3 Layer Partitioning and O1 Matrix Sharding }
\label{tab:combined_single_table}
\centering
\renewcommand{\arraystretch}{1.1}
\resizebox{\columnwidth}{!}{
\setlength{\tabcolsep}{3pt}

\begin{tabular}{>{\bfseries}c c c c c c >{\bfseries}c c c c c}
\toprule
\multicolumn{6}{c}{\textbf{(a) Forward and Backward Utilization}} & \multicolumn{5}{c}{\textbf{(b) LM Head Shard Info}} \\
\cmidrule(r){1-6} \cmidrule(l){7-11}
HS & Forward/\% & Ratio & Backward/\% & Ratio &  & HS & Shard & Section & PMU & PCU \\
\midrule
480   & 55\%  & 0.66 & 44\%   & 1.83 &  & 3072 & 9  & 2 & 316 & 504 \\
768   & 62\%  & 0.66 & 52.5\% & 2    &  & 4096 & 9  & 2 & 316 & 504 \\
1024  & 64\%  & 0.75 & 59.5\% & 2    &  & 5120 & 26 & 2 & 340 & 402 \\
1280  & 53\%  & 1    & 60.5\% & 2    &  & 6686 & 30 & 3 & 339 & 382 \\
1600  & 63\%  & 1    & 56.75\% & 3   &  & 8192 & 30 & 3 & 339 & 382 \\
\bottomrule
\end{tabular}
}
\end{table}
As HS increases in O3 mode, decoder size grows until RDU requires further partitioning. Table~\ref{tab:combined_single_table}(a) shows the "Ratio" column, which represents decoder computation sections versus total decoders. When decoders aren't further partitioned (constant ratio), PMU allocation increases with HS. However, additional partitioning causes PMU allocation to drop, explaining the oscillating pattern in O3 mode.

In O0 and O1 modes, allocation generally increases with HS until matrices require partitioning into multiple shards grouped into sections. Table~\ref{tab:combined_single_table}(b) shows PCU and PMU allocation per section correlates primarily with shard/section numbers rather than HS, suggesting resource allocation isn't tied to computational workload. PCU usage per section remains significantly below the 640 hardware limit, highlighting suboptimal scheduling.

\begin{tcolorbox}[colback=gray!20, 
  colframe=black, 
  boxsep=2pt,        
  left=2pt,          
  right=2pt,         
  top=2pt,           
  bottom=2pt         
  ]
\textbf{Insight:} RDU can train arbitrarily large models through graph partitioning and matrix sharding, but complex partitioning strategies limit resource allocation below 60\%, constraining LLM performance despite unlimited scalability.
\end{tcolorbox}

\subsection{Chip Resource Loading Balance}

Following the analysis of resource allocation ratios across different chips, we further examine whether resource assignments are balanced under platform-specific strategies. Due to the lack of IPU data, our analysis remains focused on the WSE-2 and RDU platforms.
\begin{figure} [tbp] 
	\begin{minipage}[t]{1.65in}
		\includegraphics[width=1\textwidth,height=0.12\textheight]{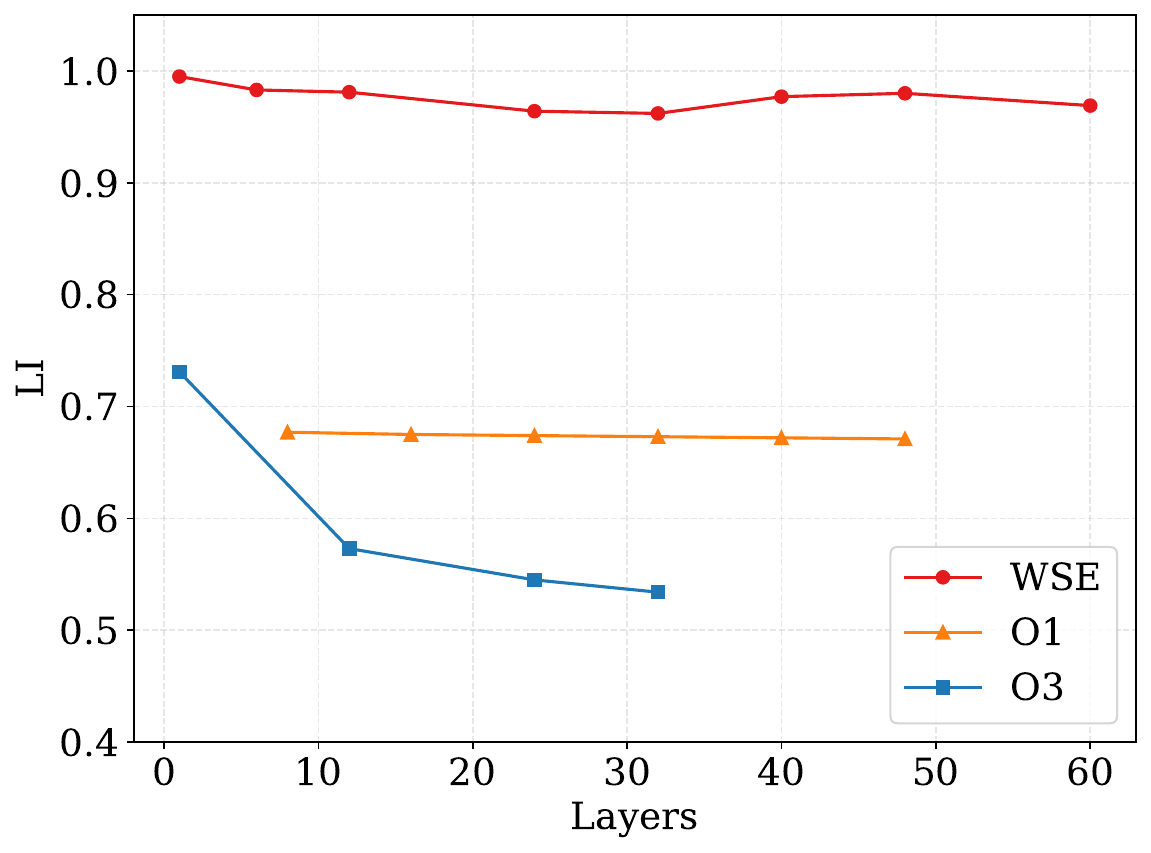}
		\par \centering (a) LI vs layer count 
	\end{minipage}
	\hfill
	\begin{minipage}[t]{1.65in}
		\includegraphics[width=1\textwidth,height=0.12\textheight]{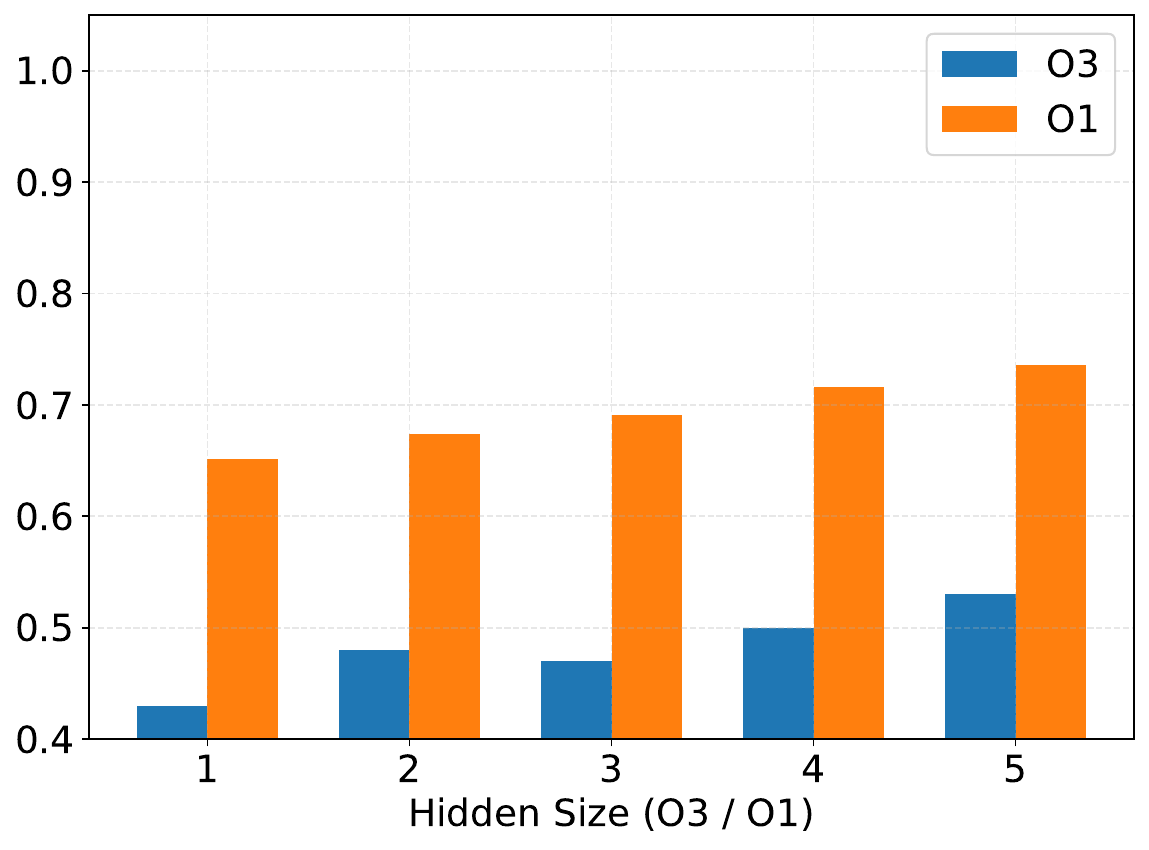}
		\par \centering (b) LI vs HS 
	\end{minipage}
	\caption{Load imbalance (LI) of WSE-2 and RDU. An LI value closer to 1 indicates better balance. 
    }
	\label{fig:hiddensize_LI}
\end{figure}
Since LI (load imbalance) is sensitive to task granularity, cross-platform comparisons are not directly meaningful; instead, LI serves as an indicator for guiding compiler-level optimization. Based on available data, we evaluate LI at the kernel level on WSE-2, and at the finer operator level on RDU.

Figure~\ref{fig:hiddensize_LI}(a) demonstrates that WSE LI values consistently remain between 0.96 and 1.0 across varying layer counts, indicating WSE-2' PE allocation strategy achieves relatively mature and effective load balance. However, these results derive from compile-time throughput estimates, suggesting actual runtime LI values may be lower. Conversely, RDU exhibits distinct patterns: O1 mode employs a single computation graph shared across decoder layers, resulting in minimal LI variation with layer count, while O3 mode shows decreasing LI as layers increase, indicating less effective load balancing across decoder layers.

Figure~\ref{fig:hiddensize_LI}(b) reveals that both O1 and O3 modes exhibit increasing LI values with growing HS, indicating improved load balance. In absolute terms, O1's operator fusion strategy demonstrates significantly superior balance compared to O3's compiler-driven automatic load strategy, achieving more balanced execution.

\begin{tcolorbox}[colback=gray!20, 
  colframe=black, 
  boxsep=2pt,        
  left=2pt,          
  right=2pt,         
  top=2pt,           
  bottom=2pt         
  ]
\textbf{Insight:} WSE-2 achieves relatively good load balance at the kernel level, while RDU presents significant optimization opportunities for operator-level load balance.
\end{tcolorbox}

\subsection{Resource Utilization Efficiency Analysis}

Beyond allocation ratio and load balance, the interaction between memory and computation plays a critical role in determining chip-level performance. This section analyzes how multilevel memory systems interact with compute resources to influence compute efficiency and expose potential performance bottlenecks.

\subsubsection{Shared Memory Level Analysis}

\begin{figure*}[htbp]
    \centering
    \begin{minipage}[t]{0.24\linewidth}
        \centering
        \includegraphics[width=\linewidth,height=0.15\textheight]{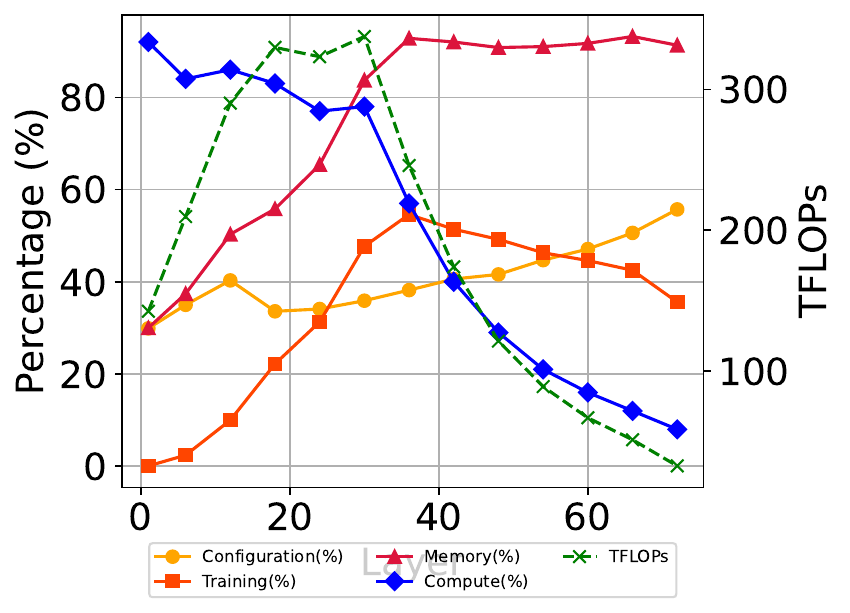}
        \par
        \vspace{2pt}
        \centering
        \footnotesize (a) WSE-2 Performance
    \end{minipage}%
    \hfill
    \begin{minipage}[t]{0.24\linewidth}
        \centering
        \includegraphics[width=\linewidth,height=0.15\textheight]{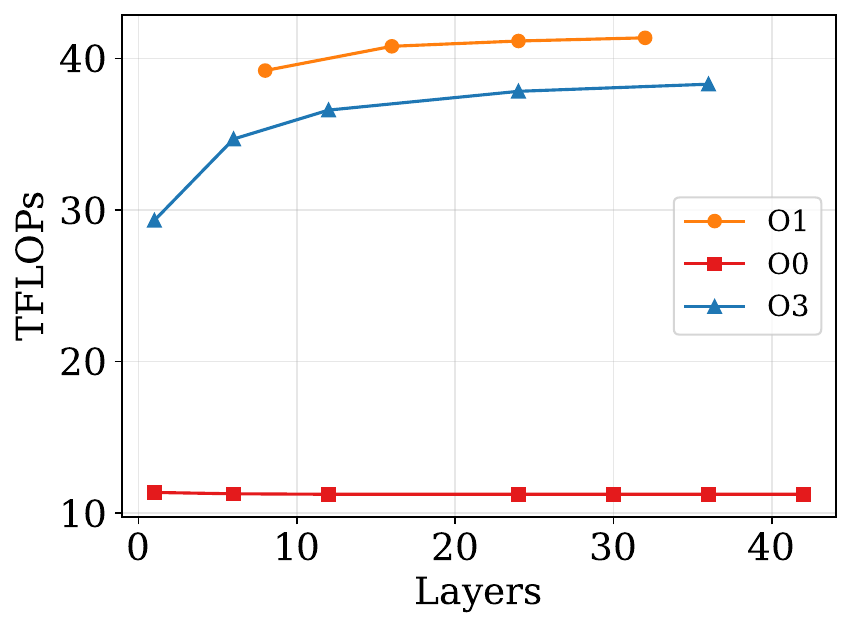}
        \par
        \vspace{2pt}
        \centering
        \footnotesize (b) RDU FLOPs vs Layers
    \end{minipage}%
    \hfill
    \begin{minipage}[t]{0.24\linewidth}
        \centering
        \includegraphics[width=\linewidth,height=0.15\textheight]{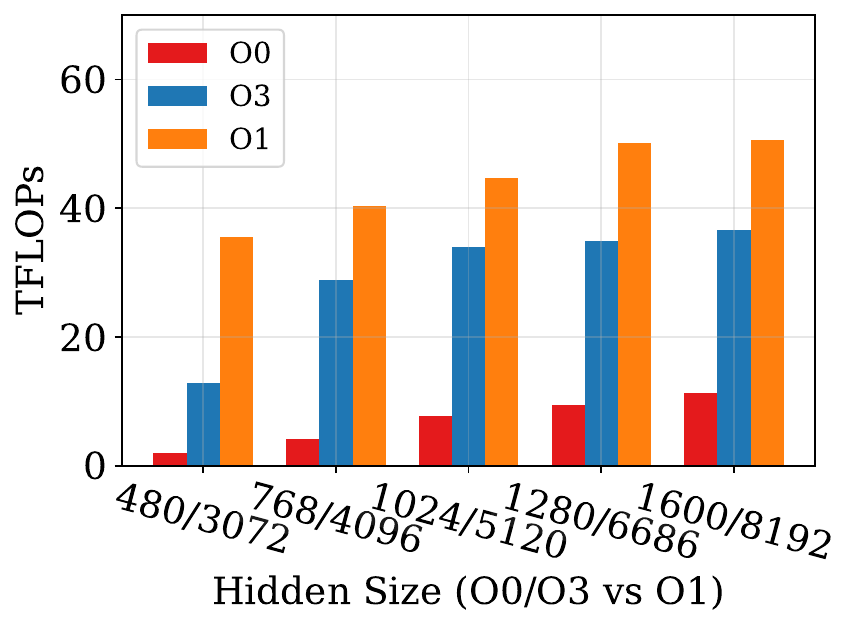}
        \par
        \vspace{2pt}
        \centering
        \footnotesize (c) RDU FLOPs by HS
    \end{minipage}%
    \hfill
    \begin{minipage}[t]{0.24\linewidth}
        \centering
        \includegraphics[width=\linewidth,height=0.15\textheight]{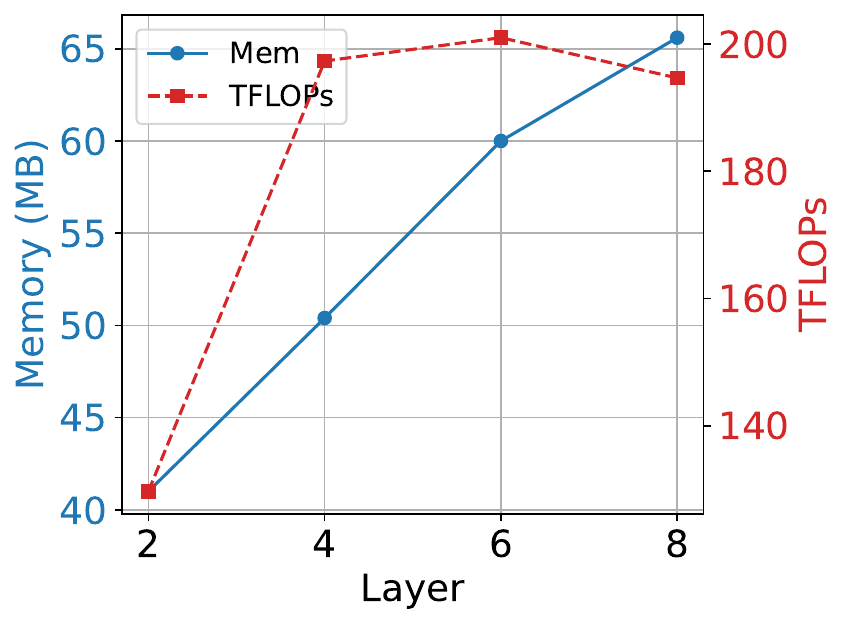}
        \par
        \vspace{2pt}
        \centering
        \footnotesize (d) IPU Performance
    \end{minipage}
    \caption{Memory utilization and compute performance analysis across chips. Figure (a) : Shows percentage breakdown versus number of layers in WSE.Each memory metric represents the proportion of configuration memory (blue), training memory (orange), and their total combined usage (red) relative to the total memory. Compute utilization (green)refers to the proportion of total runtime spent on PE computation. Figure (b) and (c): Display TFLOPS performance across different modes in RDU - O0 severely limited, while O1 and O3 show increasing performance with model scale. Figure (d): Memory usage (left y-axis) and TFLOPS (right y-axis) versus layers in IPU.}
    \label{fig:tflops_all}
\end{figure*}

These accelerators’ on-chip memories function similarly to GPU shared memory. Due to limited public data on memory bandwidth, we analyze the relationship between TFLOPs and on-chip memory usage. WSE-2 adopts a unified memory model serving both global and local roles. For RDU, due to the lack of accessible PMU utilization metrics, we approximate memory usage based on PMU allocation ratios.

\textbf{WSE-2}: From Figure \ref{fig:tflops_all}(a), we observe that TFLOPs increase steadily with the number of parameters up to 18 layers. Taken together with Table \ref{tab:pe_usage}, which shows increasing PE allocation rates and decreasing compute time per PE during this range, we conclude that when the model is small, WSE-2 performance is limited by inter-PE communication overhead, with resource allocation being the primary bottleneck.

Between 18 and 36 layers, the PE allocation rate peaks and TFLOPs stabilize, indicating efficient resource utilization. However, beyond 36 layers, memory consumption by configuration data increases sharply, reducing the memory available for training. This leads to a rapid drop in compute time (as shown by the compute metric), causing a steep decline in TFLOPs.

\textbf{RDU}:
In Figure \ref{fig:tflops_all}(b) and Figure \ref{fig:tflops_all}(c), O0 mode yields low TFLOPs due to low resource allocation. In contrast, under O1 and O3 modes, TFLOPs increase with the number of layers and HS, but the growth gradually slows down. This trend aligns with the changes in PCU allocation ratio observed in Figure \ref{fig:pmu_uti} for both modes, suggesting that variations in PMU allocation do not impose significant constraints on computational performance. The RDU can dynamically adapt the number of PMUs assigned based on the actual compute demand of the PCU, achieving balanced resource utilization.

\textbf{IPU}:
Figure \ref{fig:tflops_all}(d) shows TFLOPs increase as the number of layers grows up to 4, then plateau, and eventually fail at 10 layers (around 70M parameters). Meanwhile, memory usage increases linearly with layer count. 

\begin{tcolorbox}[colback=gray!20, 
  colframe=black, 
  boxsep=2pt,        
  left=2pt,          
  right=2pt,         
  top=2pt,           
  bottom=2pt         
  ]
\textbf{Insight}:  On WSE-2, small models are constrained by allocation inefficiency, while larger models suffer from limited on-chip memory capacity. The RDU dynamically adjusts PMU allocation based on the compute demand of each PCU, enabling balanced resource utilization. For the IPU, insufficient tile utilization limits performance at small parameter sizes, whereas the absence of flexible memory management causes the system to reach memory capacity limits at larger scales, ultimately leading to execution failure.
\end{tcolorbox}

\subsubsection{Global Memory Performance Analysis}
\begin{figure}[htbp]
    \centering
    \begin{minipage}[t]{0.33\linewidth}
        \centering
        \includegraphics[width=\linewidth,height=0.10\textheight]{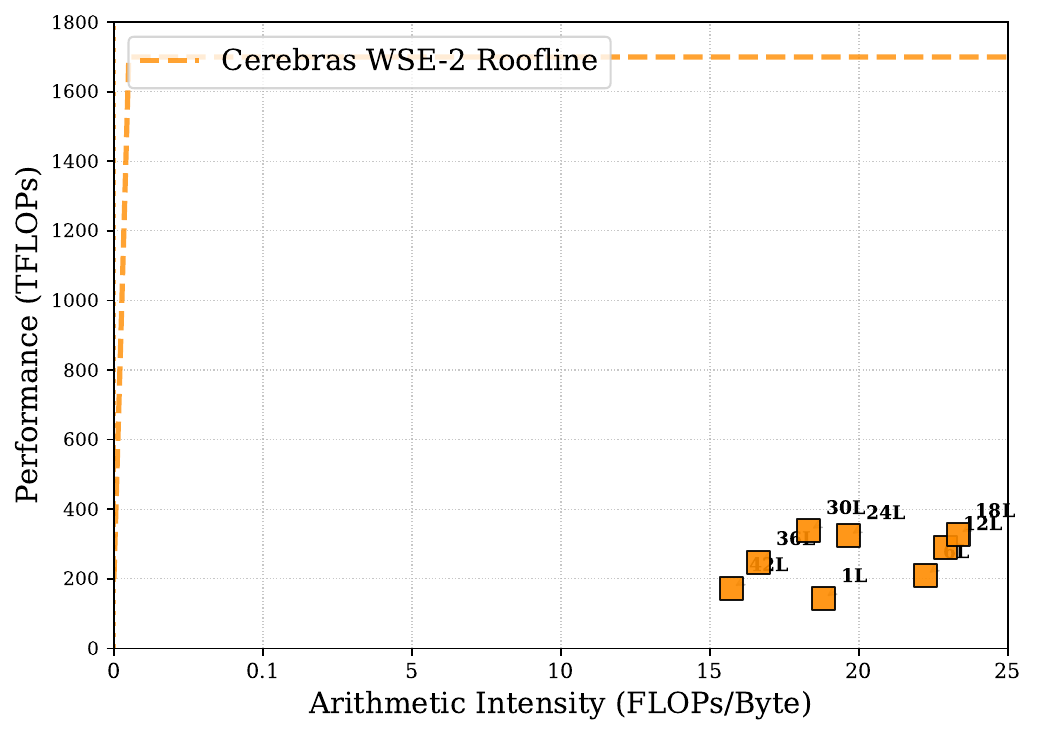}
        \par
        \vspace{2pt}
        \centering
        \footnotesize (a) Cerebras WSE-2
    \end{minipage}%
    \hfill
    \begin{minipage}[t]{0.33\linewidth}
        \centering
        \includegraphics[width=\linewidth,height=0.10\textheight]{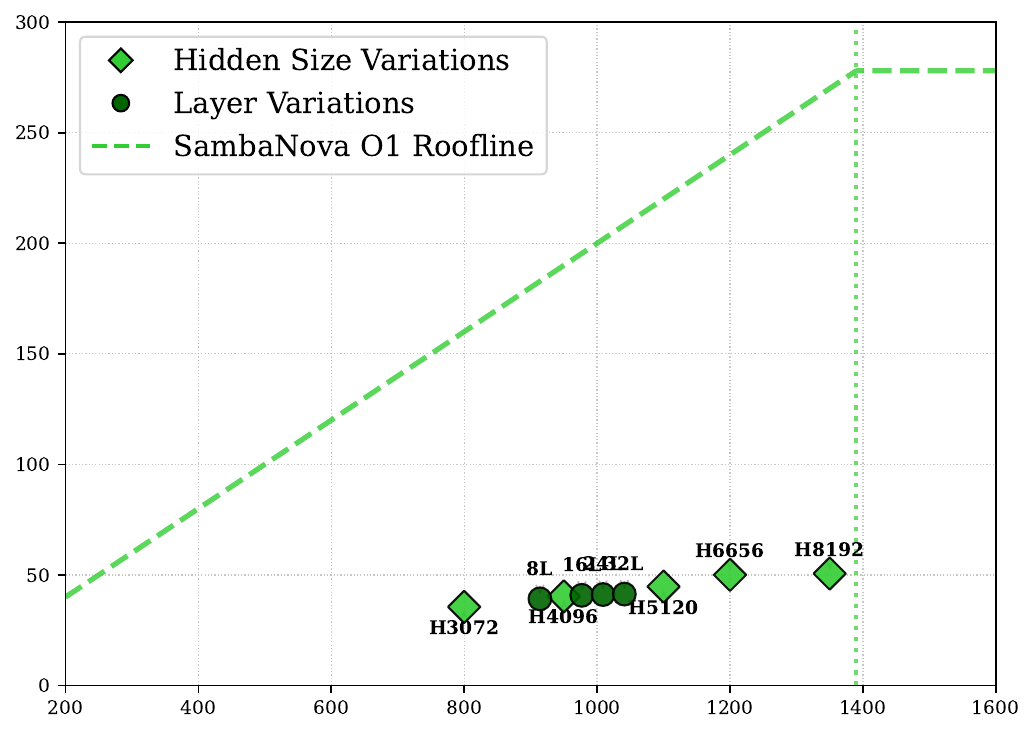}
        \par
        \vspace{2pt}
        \centering
        \footnotesize (b) SambaNova RDU
    \end{minipage}%
    \hfill
    \begin{minipage}[t]{0.33\linewidth}
        \centering
        \includegraphics[width=\linewidth,height=0.10\textheight]{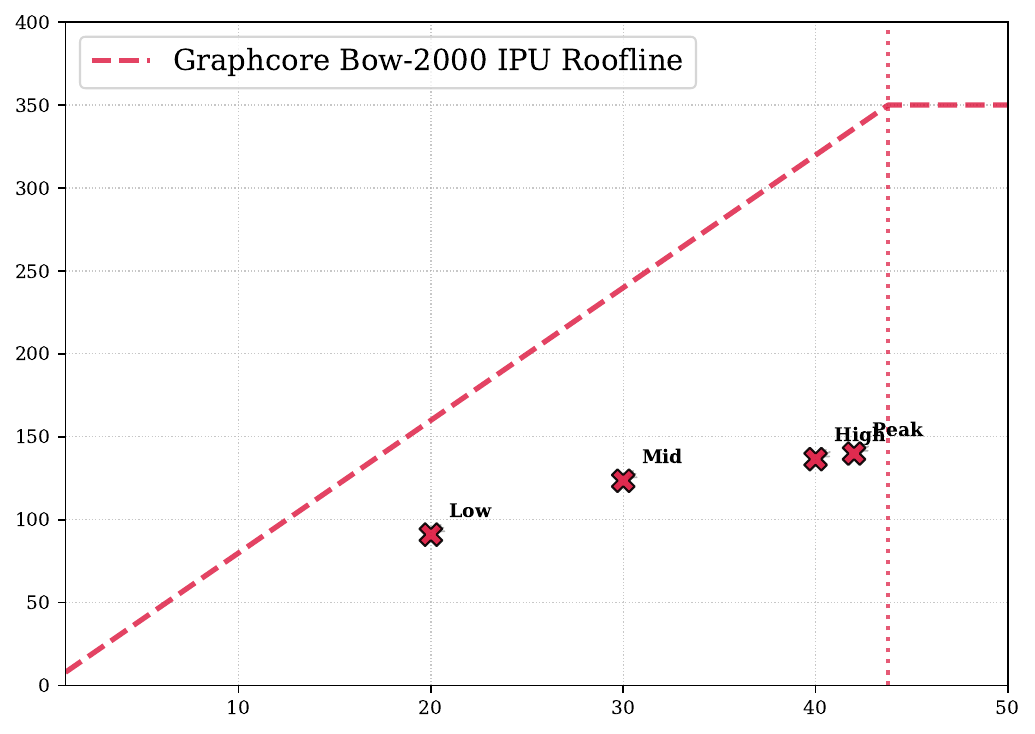}
        \par
        \vspace{2pt}
        \centering
        \footnotesize (c) Graphcore IPU
    \end{minipage}
    \caption{Roofline models across different chips. Figure (a) All workloads on WSE-2 operate in the compute-bound region.
Figure(b) and Figure (c) All workloads on RDU and IPU are memory-bound.}
    \label{fig:roofline}
\end{figure}

AI accelerator platforms adopt distinct global-memory architectures, with WSE using on-chip memory as both shared and global memory, the RDU using SN30 DDR for global memory, and the IPU using Bow-2000 DDR. Based on publicly available bandwidth specifications, we construct roofline models characterizing performance at the global memory level as shown in Figure~\ref{fig:roofline}, revealing three distinct operational paradigms.

\textbf{WSE-2:} In Figure~\ref{fig:roofline}, the exceptionally high memory bandwidth of up to 20 PB/s ensures that all workloads with arithmetic intensities in the range of 8.9–28.0 FLOPs/Byte remain compute-bound. The system achieves peak performance of 327–338 TFLOPs at 18–30 layer configurations, with compute efficiency reaching approximately 20\%. Note:This estimation is not strictly based on Equation 4, but is approximated using partial data available at compile time.

\textbf{RDU:} Limited external bandwidth (only 0.2 TB/s) causes the system to exhibit memory-bound behavior when handling LLM workloads. However, as the parameter size increases—particularly through expanding the HS—the arithmetic intensity improves, and system throughput rises from 35.55 to 50.64 TFLOPs. In contrast, increasing the number of layers yields only marginal gains, with throughput remaining in the range of 39–41 TFLOPs. The system achieves a peak compute efficiency of 18.2\%.

\textbf{IPU:} With arithmetic intensities spanning 20-42 FLOPs/Byte, most configurations operate near the memory-compute boundary. As parameter count increases, performance shows high sensitivity to workload characteristics, achieving 91-143 TFLOPs with peak efficiency of 41\% (Figure~\ref{fig:roofline}).

\begin{tcolorbox}[colback=gray!20, 
  colframe=black, 
  boxsep=2pt,        
  left=2pt,          
  right=2pt,         
  top=2pt,           
  bottom=2pt         
  ]
\textbf{Insight}: Only WSE is able to maintain a compute-bound state when running LLM workloads, primarily due to its specialized on-chip memory architecture. In contrast, both IPU and RDU exhibit typical memory-bound behavior, constrained by limited memory bandwidth. This highlights memory bandwidth as the primary bottleneck for most AI accelerators when handling LLM workloads.
\end{tcolorbox}

\textit{Conclusion.} These comprehensive Tier 1 results confirm the effectiveness of our framework in characterizing dataflow accelerators through standardized metrics. Using three key metrics, we transform previously opaque accelerator platforms into interpretable systems and reveal platform-specific bottlenecks, highlighting the framework’s practicality and generalizability.

\textit{Discussion.} Although WSE-2 demonstrates excellent PE allocation and load balancing, along with exceptionally high memory bandwidth, its compute efficiency shows no clear advantage, indicating that there is still room for optimization at the kernel level. RDU faces bottlenecks in resource utilization and bandwidth, but achieves comparable efficiency to WSE-2 through effective kernel-level optimization. Future improvements should focus on enhancing resource allocation, improving load balance, and expanding external bandwidth to unlock additional performance. IPU delivers the highest compute efficiency among the three platforms, yet all workloads remain in the memory-bound regime. Its performance is constrained by limited bandwidth and memory management capabilities. It is also recommended to improve bandwidth and incorporate strategies such as segmented loading and tensor swapping.

\section{Inter-Chip Scalability and Optimization}
This section performs Tier 2 benchmarking of our framework and evaluates the multi-chip scalability and optimization characteristics of targeted architectures, focusing on scaling strategies, bottlenecks, and the impact of batch size and memory optimizations. Results highlight distinct platform behaviors, offering guidance for hardware selection and configuration in large-scale model deployment.

\begin{table*}[ht]
\centering
\caption{\footnotesize{Multi-Hardware Scalability Performance Comparison}}
\label{tab:multi_hardware_scalability}
\resizebox{\textwidth}{!}{%
\setlength{\tabcolsep}{3pt}
\begin{tabular}{l|ccccc|cccccccc|ccc|cccccc}
\toprule
Device & \multicolumn{5}{c|}{WSE-2} & \multicolumn{8}{c|}{IPU} & \multicolumn{3}{c|}{RDU} & \multicolumn{6}{c}{GPU (Reference)} \\
\midrule 
Configuration & DP0 & DP2 & DP4 & DP8 & PP & 4PP & 4PP & 8PP & 8PP & 16PP & 16PP & 16PP & 16PP & TP2 & TP4 & TP8 & T8P1D1 & T4P2D1 & T2P4D1 & T1P8D1 & T8P8D16 & T4P4D64 \\
Model & small & small & mini & tiny & small & 6L & 12L & 18L & 24L & 30L & 36L & 42L & 48L & 7B & 7B & 7B & xlarge & xlarge & xlarge & xlarge & xlarge & xlarge \\
Throughput & 0.66M & 0.98M & 1.84M & 3.6M & 0.53M & 120 & 80 & 129 & 105.4 & 223 & 181 & 178 & 153 & 1540 & 945 & 918 & 155.3 & 145.2 & 135.8 & 120.4 & 163.2 & 158.9 \\
\bottomrule
\end{tabular}
}
\end{table*}

\subsection{Scalability Benchmarking}

\subsubsection{Scaling Strategies}
Different accelerators adopt parallel strategies inspired by GPU paradigms: IPUs implement pipeline parallelism (PP) by assigning model layers to different devices; SambaNova achieves tensor parallelism (TP) by partitioning operators across RDUs. WSE-2 is distinct in that its scalability does not rely on multi-chip parallelism. Instead, it leverages intra-chip data parallelism (DP) and a weight streaming mode resembling PP, all within a single large chip. Since PP on CS-2 requires root access, we only collected limited PP data on CS-3, and further exploration was constrained by platform limitations. Other strategies were not adopted due to immature software support or possible oversight.

\subsubsection{Performance Overview}

Table~\ref{tab:multi_hardware_scalability} presents comprehensive scalability performance across all evaluated platforms. GPU results are included as reference baselines for comparison with specialized dataflow accelerators. 
\subsubsection{Insights into Scalabilities}

\begin{figure}[htbp]
    \centering
    \begin{minipage}[t]{0.33\linewidth}
        \centering
        \includegraphics[width=\linewidth,height=0.09\textheight]{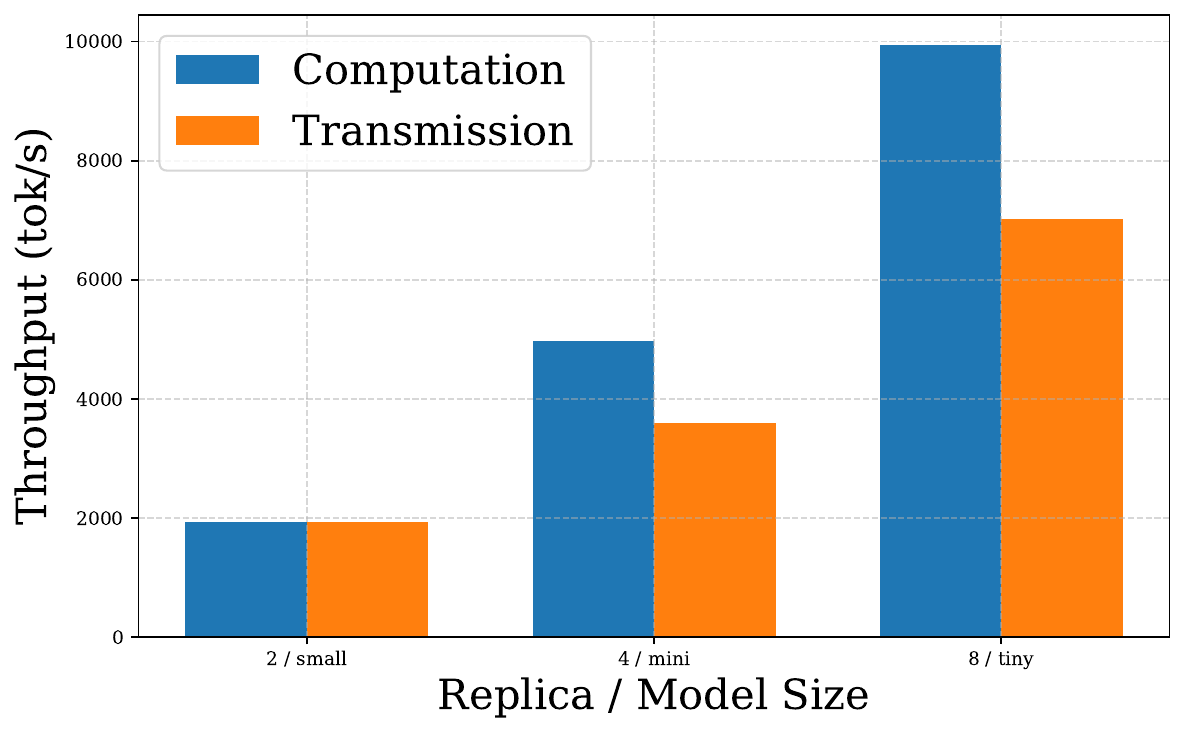}
        \par
        \vspace{2pt}
        \centering
        \footnotesize (a) WSE Throughput vs Replica
    \end{minipage}%
    \hfill
    \begin{minipage}[t]{0.33\linewidth}
        \centering
        \includegraphics[width=\linewidth,height=0.09\textheight]{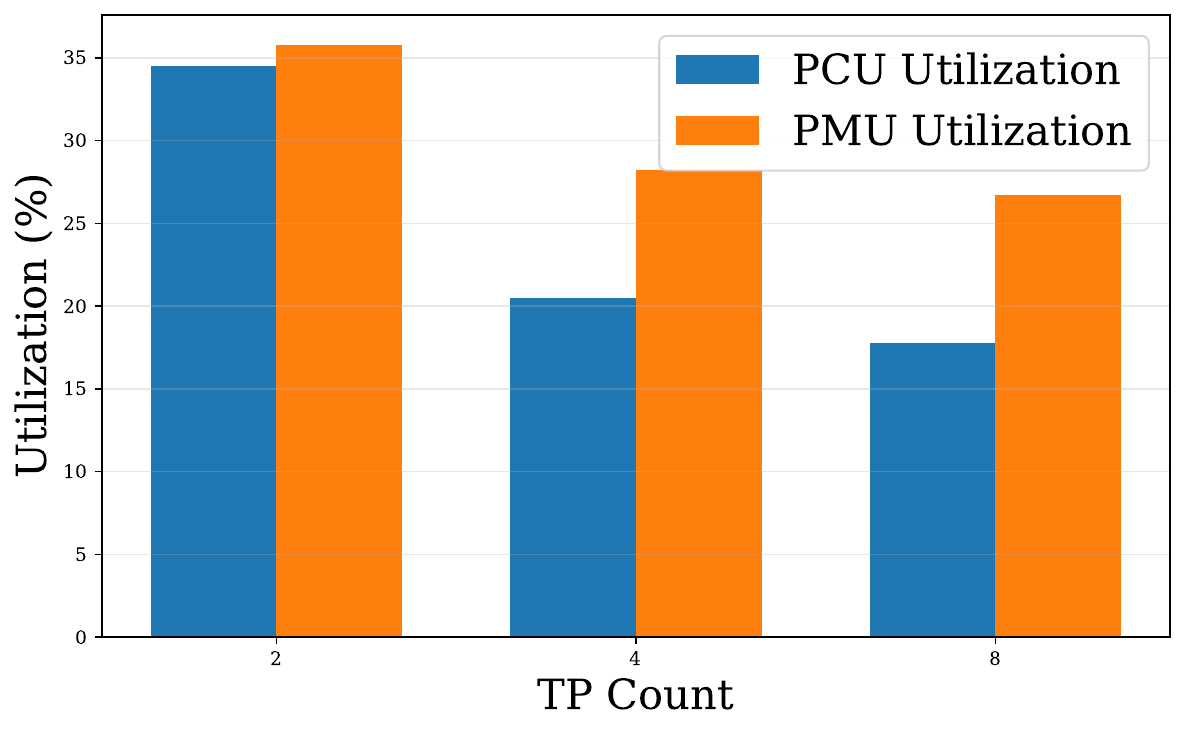}
        \par
        \vspace{2pt}
        \centering
        \footnotesize (b) RDU Utilization vs TP
    \end{minipage}%
    \hfill
    \begin{minipage}[t]{0.33\linewidth}
        \centering
        \includegraphics[width=\linewidth,height=0.09\textheight]{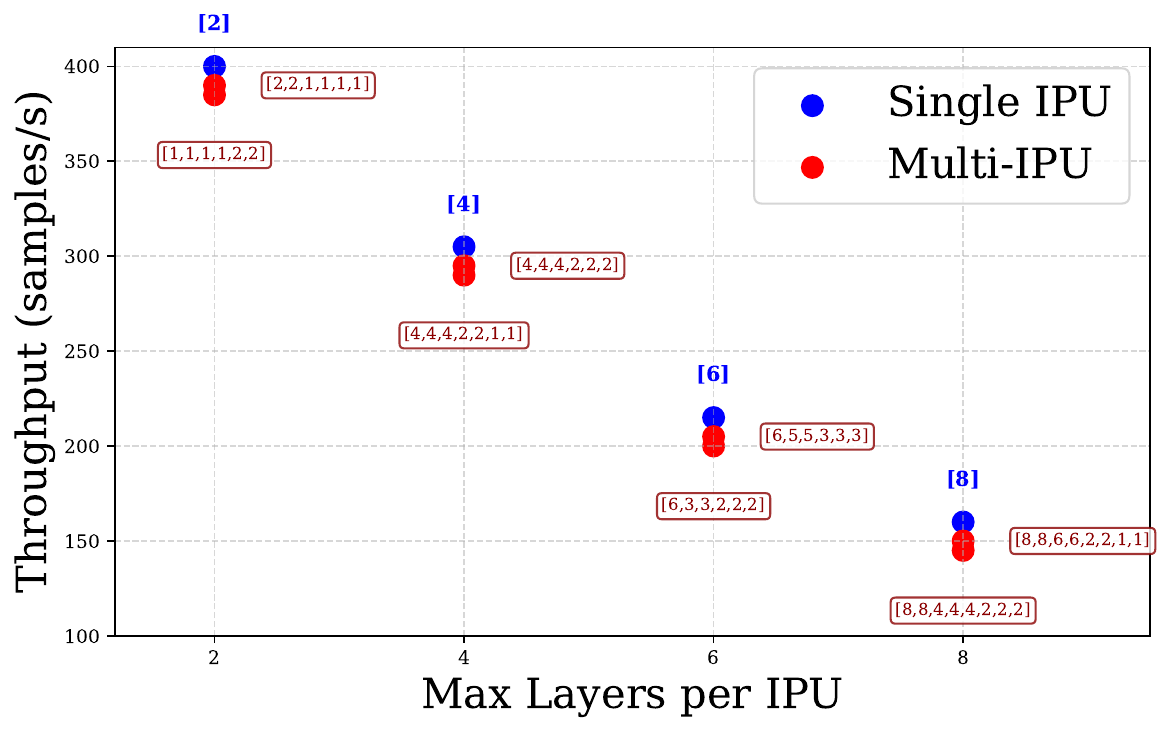}
        \par
        \vspace{2pt}
        \centering
        \footnotesize (c) IPU Throughput vs Layers
    \end{minipage}
    \caption{Scalability details across dataflow hardware: Figure (a) shows WSE throughput under varying replica counts, where the gap between computation and communication reflects overhead; Figure (b) presents average RDU resource utilization across different TP configurations; Figure (c) illustrates IPU throughput under various layer allocations, highlighting that performance is limited by the most heavily loaded IPU.}
    \label{fig:scalability_detail}
\end{figure}

\paragraph{Cerebras WSE Scalability} Based on its architectural characteristics, the WSE offers two distinct scalability modes. The first mode applies to relatively small models. To utilize more PEs, the WSE-2 supports intra-chip data parallelism, which creates multiple data-parallel replicas—full copies of the model running independently on separate WSE-2 partitions. As shown in Figure \ref{fig:scalability_detail} (a), smaller models support a greater number of parallel replicas. The more replicas are deployed, and the higher the communication cost relative to computation. With only two replicas, optimized placement can reduce the communication distance between them to zero by ensuring adjacency of communication paths. For larger numbers of replicas, however, such optimal layouts are no longer achievable.

When the WSE-2 chip cannot accommodate the entire model, it switches to weight streaming mode. In this mode, only a portion of the model is loaded onto the chip at a time: the outputs from the previous layer are retained on-chip, while the weights for the next layer are streamed in for computation. As shown in Table \ref{tab:multi_hardware_scalability}, the throughput of GPT-2 decreases from 0.66M tokens/s to 0.53M tokens/s under this mode—a reduction of only about 20\%. 

\begin{tcolorbox}[colback=gray!20, 
  colframe=black, 
  boxsep=2pt,        
  left=2pt,          
  right=2pt,         
  top=2pt,           
  bottom=2pt         
  ]
\textbf{Insight}: On WSE-2, applying intra-chip data parallelism to smaller models can significantly improve compute performance. However, as the number of replicas increases, the total on-chip communication distance grows, making communication overhead a performance bottleneck. For larger models, the weight streaming mode can be employed, which introduces only modest additional overhead.
\end{tcolorbox}

\paragraph{SambaNova RDU Scalability}

As shown in Table \ref{tab:multi_hardware_scalability}, experiments on SambaNova’s TP reveal that scaling from 2 to 4 chips results in a ~40\% drop in throughput (from 1540 to 945 tokens/s), while further scaling introduces minimal additional overhead. Since each SN30 device contains two RDUs, this indicates that intra-machine TP incurs low communication cost, whereas inter-machine TP becomes the dominant bottleneck. As shown in Figure \ref{fig:scalability_detail}(b), Resource allocation data further supports this: cross-machine TP reduces PCU and PMU allocation per chip by approximately 40\% and 25\%, respectively, indicating that communication-heavy sections significantly increase, leading to severe underutilization of chip resources. 
\begin{tcolorbox}[colback=gray!20, 
  colframe=black, 
  boxsep=2pt,        
  left=2pt,          
  right=2pt,         
  top=2pt,           
  bottom=2pt         
  ]
\textbf{Insight}: On RDU, tensor parallelism addresses DDR memory limits for large models, but cross-machine TP incurs high communication overhead. It should be avoided when single-machine memory is sufficient to prevent performance degradation.
\end{tcolorbox}

\paragraph{Graphcore IPU Scalability}

Figure \ref{fig:scalability_detail}(c) presents the throughput results under nine different load distribution configurations. The experiments show that when decoder layers are distributed across multiple IPUs, overall system throughput is primarily limited by the most heavily loaded IPU, indicating that performance bottlenecks are concentrated on the device assigned the most work.

Table \ref{tab:multi_hardware_scalability} reports experimental data across different layer configurations. Taken together with the observation from Figure \ref{fig:tflops_all}(d)—that IPUs quickly reach peak TFLOPs even for small models—it becomes evident that, once TFLOPs stabilize, an IPU’s throughput is roughly inversely proportional to the number of layers it handles.

\begin{tcolorbox}[colback=gray!20, 
  colframe=black, 
  boxsep=2pt,        
  left=2pt,          
  right=2pt,         
  top=2pt,           
  bottom=2pt         
  ]
\textbf{Insight}: When using PP on IPUs, overall throughput is determined by the maximum number of layers assigned to any single IPU. Deployment should minimize this maximum. If the most heavily loaded IPU’s compute resources are fully utilized, throughput degradation from load imbalance remains minimal.
\end{tcolorbox}

\subsection{Deployment Optimization}

This section evaluates the impact of batch size and different precision formats on different accelerators during LLM deployment. 
\begin{figure}[htbp]
\centering
\begin{minipage}[t]{0.32\linewidth}
\centering
\includegraphics[width=\linewidth]{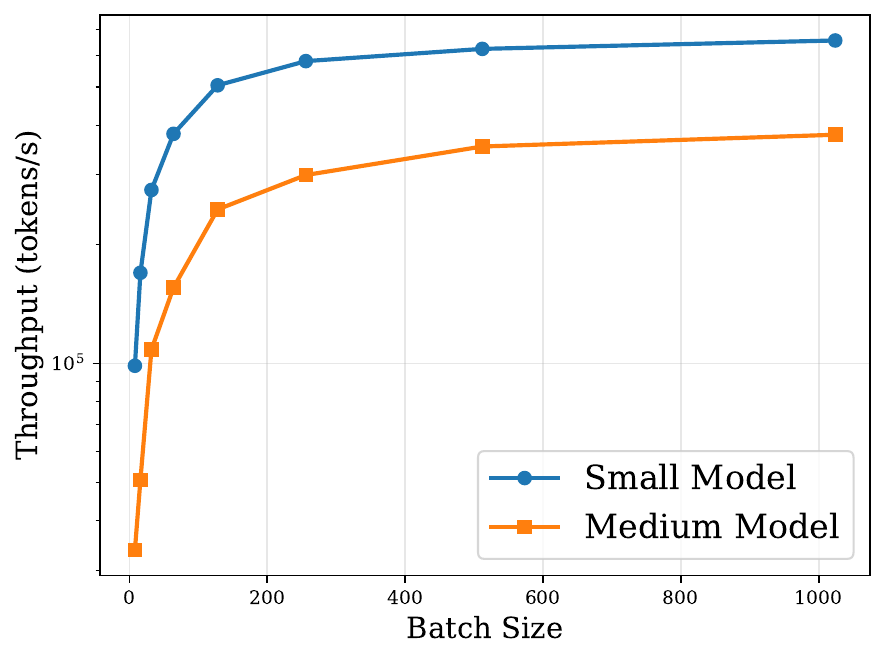}
\par
\vspace{2pt}
\centering
\footnotesize (a) WSE Platform
\end{minipage}%
\hfill
\begin{minipage}[t]{0.32\linewidth}
\centering
\includegraphics[width=\linewidth]{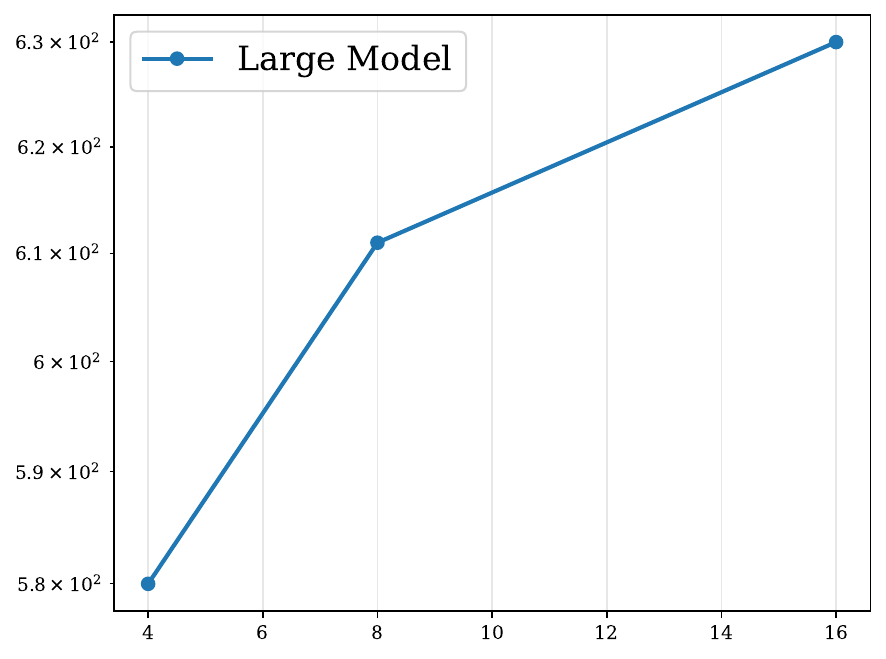}
\par
\vspace{2pt}
\centering
\footnotesize (b) RDU Platform
\end{minipage}%
\hfill
\begin{minipage}[t]{0.32\linewidth}
\centering
\includegraphics[width=\linewidth]{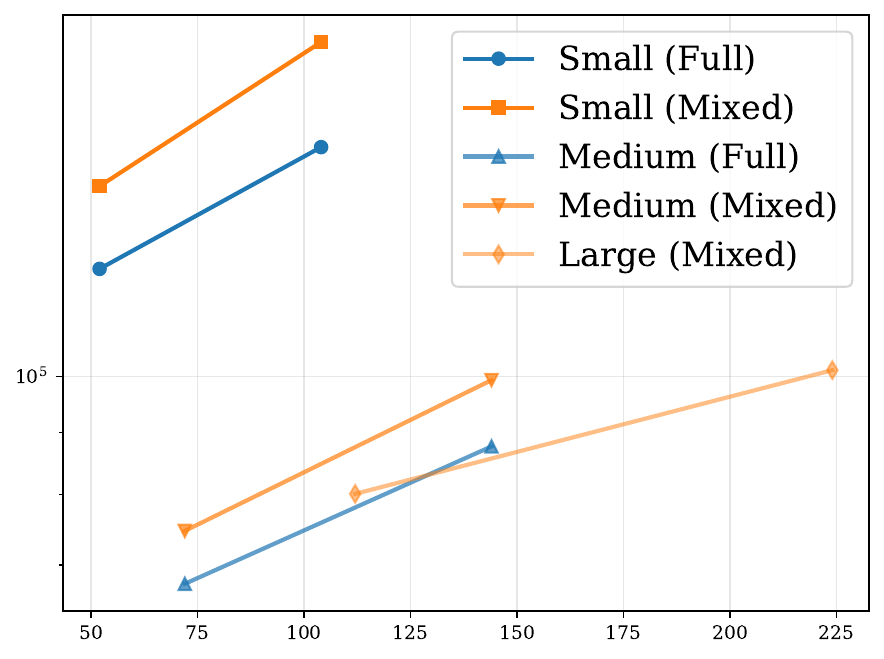}
\par
\vspace{2pt}
\centering
\footnotesize (c) IPU Platform
\end{minipage}
\caption{Throughput performance analysis across different hardware platforms showing batch size scaling behavior.}
\label{fig:throughput_analysis}
\end{figure}

\paragraph{Batch Size Impact Analysis}
As shown in the Figure \ref{fig:throughput_analysis}, increasing the batch size on both IPU and RDU generally yields a near-linear improvement in throughput. In contrast, on the WSE, we observe a significant gain in throughput when the batch size is below 200, but the benefit becomes less pronounced beyond that point. Therefore, we recommend using a batch size greater than 200 on the WSE, while maximizing the batch size as much as possible on other platforms.

\begin{table}[htbp]
\small
\centering
\caption{Mixed Precision Throughput Across Platforms}
\label{tab:mixed_precision_split_columns}
\resizebox{\linewidth}{!}{%
\begin{tabular}{@{}lcccccc@{}}
\toprule
\textbf{Device} & \multicolumn{2}{c}{\textbf{IPU}} & \multicolumn{2}{c}{\textbf{WSE}} & \multicolumn{2}{c}{\textbf{RDU}}(7B) \\
\cmidrule(lr){2-3} \cmidrule(lr){4-5} \cmidrule(lr){6-7}
\textbf{Configuration} & Full & Mixed & FP16 & CB16 & BF16 & Mixed \\
\midrule
\textbf{Throughput} & 154k & 188k & 527k & 583k & 631 & 847 \\
\bottomrule
\end{tabular}%
}
\label{tab:mix}
\end{table}

\paragraph{Different Precision Analysis}

As shown in Table \ref{tab:mix}, different accelerators support different precision formats, and the impact of precision on throughput varies significantly. RDU is the most sensitive to precision, with mixed-precision training yielding a 34.3\% throughput improvement. IPU follows, with a gain of 22.0\%, while WSE shows minimal impact, with only a 10.7\% increase. Therefore, precision selection should be carefully considered when deploying models on RDU and IPU, whereas for WSE, the effect is relatively limited.


\begin{tcolorbox}[colback=gray!20, 
  colframe=black, 
  boxsep=2pt,        
  left=2pt,          
  right=2pt,         
  top=2pt,           
  bottom=2pt         
  ]
\textbf{Insight}: When deploying models, use the largest possible batch size on RDU and IPU to maximize throughput. On WSE, avoid batch sizes below 200 due to performance drops.
Precision options are limited across accelerators, but RDU and IPU benefit significantly from mixed precision, while WSE shows minimal sensitivity.
\end{tcolorbox}

\textit{Conclusion.} This section systematically evaluates inter-chip scalability and performance bottlenecks of different accelerators through DP, TP and PP. It also analyzes the impact of key deployment parameters—such as batch size and precision—on throughput across platforms, thereby validating the effectiveness and generalizability of Tier 2 in guiding optimized LLM deployment strategies.

\textit{ Discussion.} WSE-2’s scalability is primarily based on designing different execution modes for a single chip. Beyond the recommendations in Section 5, further optimization should focus on on-chip kernel placement and external bandwidth provisioning to fully unlock its computational potential. In contrast, enhancing the scalability of RDU should focus on two key aspects: expanding external memory capacity to reduce cross-machine communication, and minimizing communication overhead through techniques such as overlapping communication with computation and increasing memory bandwidth. Meanwhile, due to its limited memory capacity, the IPU requires further investigation into effective parallelization strategies for handling language models with large decoders. In addition, when deploying these models, it is recommended to use sufficiently large batch sizes and lower precision to maximize acceleration.
\section{Related work}
\textbf{Applications of AI Accelerators.} A growing number of studies have focused on porting specific algorithms or applications to AI accelerator platforms to improve computational efficiency\cite{brown2020language,hoffmann2024exploring,trotter2021epigenomic,acar2025optimizing}.\cite{moe2022implementing} deploys a spatiotemporal graph convolutional network on the Graphcore IPU to accelerate video action recognition tasks. In \cite{louw2021using}, several representative scientific computing tasks—such as linear algebra operations and Conjugate Gradient methods—are implemented on the IPU. \cite{ltaief2023scaling} applies a Tile Low-Rank algebraic compression method on the Cerebras CS-2 to accelerate multidimensional deconvolution in seismic data processing. In \cite{song2024ceresz}, an error-bounded lossy compression algorithm is implemented on CS-2, achieving significant performance gains over traditional GPU-based solutions while maintaining accuracy.

\textbf{Existing Benchmarking on AI Accelerators.} Existing benchmarks primarily focus on throughput performance while neglecting the characterization of internal architectural behaviors of AI accelerators. One line of research focuses on evaluating how AI accelerators perform on domain-specific scientific applications. For instance, \cite{arcelin2021comparison} and \cite{mohan2020studying} benchmark Graphcore IPUs on commonly used methods in particle physics and astrophysics to assess their acceleration potential in those fields. Another line of work emphasizes benchmarking various AI accelerators for standard AI workloads. \cite{emani2022comprehensive, emani2024toward} benchmark AI accelerators by measuring throughput on neural network primitives (e.g., ReLU, convolution) and large language models (BERT, GPT), with additional analysis of parameter sensitivity. In addition, \cite{peng2024evaluating} evaluates the performance of IPUs and RDUs on dense (GEMM) and sparse (SPMM) matrix operations. 

\section{Conclusion}

This paper introduces DABench-LLM, the first benchmarking framework tailored to assess AI accelerators with dataflow architectures under LLM workloads. Validated on Cerebras CS-2, SambaNova SN30, and Graphcore Bow-2000, it systematically uncovers their resource scheduling policies and execution patterns. By transforming these platforms from opaque systems into transparent, interpretable architectures, the framework offers critical insights into their design trade-offs and performance characteristics. It also lays the groundwork for future research on bottleneck identification and platform-specific optimization.

\section*{Acknowledgment}

This research used resources of the Argonne Leadership Computing Facility, a U.S. Department of Energy (DOE) Office of Science user facility at Argonne National Laboratory, and is based on research supported by the U.S. DOE Office of Science-Advanced Scientific Computing Research Program, under Contract No. DE-AC02-06CH11357. 
This research used resources of the Advanced Photon Source, a U.S.\ DOE Office of Science user facility at Argonne National Laboratory, and is based on research supported by the DOE Office of Science-Basic Energy Sciences, under Contract No. DE-AC02-06CH11357. 
This work also used the Cerebras CS-2 on the Pittsburgh Supercomputing Center's (PSC) Neocortex system~\cite{buitrago2021neocortex}.

We thank Zhuo Chen and Luke Jiang of SambaNova Systems, and Julian Uran, Leighton Wilson, and Mei-Yu Wang of the Pittsburgh Supercomputing Center, for their helpful feedback and guidance.
\bibliographystyle{IEEEtranS}
\bibliography{reference}
\clearpage
\newpage

\appendix
\subsection{Abstract}
This artifact provides a comprehensive collection of test scripts designed to evaluate performance across hardware configurations, including Graphcore IPU, Cerebras, and SambaNova systems. It includes references to key articles and papers that informed the development and validation of these tests, enabling reproducibility of the experiments described in the work \textit{DABench-LLM: Standardized and In-Depth Benchmarking of Post-Moore Dataflow AI Accelerators for LLMs}. The artifacts consist of source code, datasets, scripts, and documentation to replicate the benchmarking results for LLMs on the targeted dataflow accelerators. These accelerators are available through the ALCF AI Testbed (https://www.alcf.anl.gov/alcf-ai-testbed), but access is restricted, and the authors are not authorized to grant it.

\subsection{Artifact check-list (meta-information)}
{\small
\begin{itemize}
  \item {\bf Algorithm: } Benchmarking algorithms for LLMs on dataflow accelerators
  \item {\bf Model: } LLMs sourced from official GitHub repositories (Graphcore examples, SambaNova tutorials, Cerebras modelzoo)
  \item {\bf Data set: } Datasets downloaded via scripts (e.g., for LLM benchmarking)
  \item {\bf Run-time environment: } ALCF AI Testbed (Linux-based, Python 3.10+)
  \item {\bf Hardware: } Graphcore Bow Pod64, Cerebras CS-2, SambaNova DataScale SN30
  \item {\bf Run-time state: } Managed via Slurm on ALCF nodes
  \item {\bf Execution: } Shell scripts and Python commands
  \item {\bf Metrics: } Performance metrics (e.g., throughput, scaling efficiency)
  \item {\bf Output: } Analysis logs and results from ana.py
  \item {\bf Experiments: } Benchmark scaling runs on each accelerator.
  \item {\bf How much disk space required: } 50 GB (including datasets and models)
  \item {\bf How much time is needed to prepare workflow : } 30 minutes (account setup and cloning)
  \item {\bf How much time is needed to complete experiments: } Several hours per accelerator (varies by scale)
  \item {\bf Publicly available: } Yes
  \item {\bf Code licenses : } MIT License (assumed; check LICENSE file in repo)
  \item {\bf Data licenses : } Varies by dataset source
  \item {\bf Workflow automation framework used?: } Shell scripts and Slurm
  \item {\bf Archived: } No (available via GitHub)
\end{itemize}
}
\subsection{Description}
\subsubsection{How to access}
The artifacts are publicly available on GitHub at \url{https://github.com/augustuszzq/Regular-DABench-LLM.git}.
Clone the repository using: \texttt{git clone https://github.com/augustuszzq/Regular-DABench-LLM.git}.
\subsubsection{Hardware dependencies}
- Graphcore Bow Pod64  
- Cerebras CS-2
- SambaNova DataScale SN30
\subsubsection{Software dependencies}
- OS: Linux-based (e.g., Ubuntu 20.04 LTS or later, as managed in the ALCF environment).  
- Programming Language: Python 3.10+ (integrated with machine learning frameworks).  
- Dependencies: Listed in \texttt{requirements.txt}. Key dependencies include:  
  - Graphcore: Poplar SDK (for graph software and ML applications), PopTorch (PyTorch wrapper optimized for IPUs), PopLibs (for tensor and graph operations), and integrations with TensorFlow/PyTorch.  
  - Cerebras: PyTorch framework integration for model compilation and execution; managed as an appliance for data preprocessing, streaming, and orchestration.  
  - SambaNova: SambaFlow software stack (for optimizing and mapping dataflow graphs to RDUs), with integrations to PyTorch.
\subsubsection{Data sets}
Datasets are downloaded using the provided script: \texttt{bash scripts/download\_data.sh} or manually from official sources via wget or curl. They are stored in the \texttt{/data/} directory. For large datasets, utilize ALCF storage systems to avoid node limits.
\subsubsection{Models}
The models are derived from the following official GitHub repositories:  
- \textbf{Graphcore IPU}: Sourced from \url{https://github.com/graphcore/examples.git}. 
- \textbf{SambaNova}: Sourced from \url{https://github.com/sambanova/tutorials/tree/main}. 
- \textbf{Cerebras}: Sourced from \url{https://github.com/Cerebras/modelzoo.git}. T

These sources ensure standard, unmodified official implementations. Models are loaded by cloning the respective repositories and following their setup instructions.
\subsection{Installation}
Since this artifact relies on hardware provided by Argonne National Laboratory (ALCF) AI Testbed, installation and setup are performed within the ALCF environment. Users must have an ALCF account with Multi-Factor Authentication (MFA) enabled (e.g., via MobilePASS+ or CRYPTOCard). The process involves accessing the respective hardware nodes, cloning the repository, and configuring the environment specific to each hardware type (Graphcore IPU, Cerebras CS-2, SambaNova SN30).  

\subsubsection{General Prerequisites}  
- \textbf{ALCF Account}: Request an account via the ALCF "Get Started" guide if you don't have one. Enable MFA for secure access.  
- \textbf{SSH Access}: Use SSH from your local machine with your ALCF user ID and MFA-generated passcode.  
- \textbf{Repository Cloning}: All setups start by cloning this GitHub repository on the ALCF nodes.  
- \textbf{Dependencies}: Python-based; key packages include those for ML frameworks (e.g., PyTorch, TensorFlow). Specific versions and installations vary by hardware.  

\subsubsection{1. Access ALCF Systems}  
Access follows a two-step SSH process for most hardware:  
- From your local machine: \texttt{ssh ALCFUserID@ai.alcf.anl.gov} (replace \texttt{ALCFUserID} with your ID; enter MFA passcode as password). Use \texttt{-v} for debugging (e.g., \texttt{ssh -v ALCFUserID@ai.alcf.anl.gov}).  
- Once on the login node, SSH to the specific hardware node (details below).  

\subsubsection{2. Clone the Repository}  
On the target hardware node (after accessing it):  
\begin{verbatim}
git clone https://github.com/augustuszzq/Regular-DABench-LLM.git
cd Regular-DABench-LLM
\end{verbatim}  

\subsubsection{3. Hardware-Specific Setup and Installation}  

\paragraph{Graphcore IPU}  
- \textbf{Access Nodes}: From the login node, SSH to an accessible Graphcore node, e.g., \texttt{ssh gc-poplar-02.ai.alcf.anl.gov}, \texttt{ssh gc-poplar-03.ai.alcf.anl.gov}, or \texttt{ssh gc-poplar-04.ai.alcf.anl.gov}. Note: \texttt{gc-poplar-01.ai.alcf.anl.gov} is not directly accessible; its IPU resources are allocated via Slurm.  
- \textbf{Environment Setup}: The Poplar SDK and related tools (e.g., PopTorch for PyTorch integration, PopLibs for tensor operations) are pre-configured in the ALCF environment. No manual setup is needed beyond login.  
- \textbf{Install Dependencies}:  
  - Activate a Python virtual environment if desired:   
    \begin{verbatim}
    python -m venv env
    source env/bin/activate
    \end{verbatim}  
  - Install project-specific packages: \texttt{pip install -r requirements.txt}.  
  - Key packages: numpy==1.26.0, torch==2.0.1 (optimized via PopTorch), and Graphcore-specific libraries (pre-installed via Poplar SDK).

\paragraph{Cerebras CS-2}  
- \textbf{Access Nodes}: From the login node, SSH to a Cerebras login node (specific node names like cs2-login-01.ai.alcf.anl.gov; use random assignment if multiple). Authentication uses MFA passcode.  
- \textbf{Environment Setup}: The Cerebras software stack is managed as an appliance, including PyTorch integration for model compilation and execution. Environment variables and tools are auto-set upon login. Includes support for data preprocessing, streaming, and orchestration via MemoryX and SwarmX nodes.  
- \textbf{Install Dependencies}:  
  - Create a virtual environment:  
    \begin{verbatim}
    python -m venv env
    source env/bin/activate
    \end{verbatim}  
  - Install via: \texttt{pip install -r requirements.txt}.  
  - Key packages: torch==2.0.1 (Cerebras-optimized), numpy==1.26.0, and dependencies for wafer-scale processing (pre-integrated).  
- \textbf{Job Submission}: Submit workflows via Cerebras-specific commands or Slurm; see the ALCF workflows section for details (e.g., for training/inference on the wafer-scale cluster).  
- \textbf{Troubleshooting}: Check MFA setup; refer to ALCF support for node access issues.  

\paragraph{SambaNova SN30}  
- \textbf{Access Nodes}: From the login node, SSH to a SambaNova node using aliases like \texttt{ssh sn30-r1-h1} (format: sn30-r[1-4]-h[1-2], where r=rack, h=host).  
- \textbf{Environment Setup}: The SambaFlow software stack is automatically set up upon login, including environmental variables. SambaFlow optimizes dataflow graphs for RDUs and integrates with PyTorch.  
- \textbf{Install Dependencies}:  
  - Virtual environment (optional):  
    \begin{verbatim}
    python -m venv env
    source env/bin/activate
    \end{verbatim}  
  - Run: \texttt{pip install -r requirements.txt}.  
  - Key packages: torch==2.0.1 (SambaNova-integrated), numpy==1.26.0, and SambaFlow-specific tools (pre-installed).  
- \textbf{Job Submission}: Use standard commands for model parallelism; refer to SambaNova tutorials for running on RDUs.  
- \textbf{Troubleshooting}: Debug SSH with \texttt{-v}; consult ALCF for account issues.

\subsection{Experiment workflow}
To reproduce the experiments, follow the hardware-specific setups above. For a quick start on Graphcore:  
\begin{verbatim}
cd graphcore
./full_run_benchmark_scalling.sh
python ana.py
\end{verbatim}  

Similar workflows apply to other hardware; refer to repository directories for Cerebras and SambaNova-specific scripts. Use Slurm for job submission where required.
\subsection{Evaluation and expected results}
Running the experiments should reproduce the benchmarking results from the paper, including performance metrics such as throughput, scaling efficiency, and comparisons across hardware for LLMs. Outputs from \texttt{ana.py} will generate analysis logs and figures matching those in the paper (e.g., performance improvements or bottlenecks). Results may vary slightly due to hardware scheduling or randomness, but should align within reasonable margins. Verify by comparing generated outputs in the results directory to paper figures.
\subsection{Experiment customization}
Experiments can be customized by modifying scripts (e.g., parameters in \texttt{full\_run\_benchmark\_scalling.sh}) or config files if present. For different scales, adjust Slurm job parameters or input sizes. Refer to inline comments in scripts for details.
\subsection{Notes}
This artifact is designed for the ALCF AI Testbed environment; local reproduction may not be feasible without equivalent hardware. For questions, open an issue on GitHub. Last updated: August 21, 2025.
\subsection{Methodology}
Submission, reviewing and badging methodology:
\begin{itemize}
  \item \url{https://www.acm.org/publications/policies/artifact-review-and-badging-current}
  \item \url{https://cTuning.org/ae}
\end{itemize}

\end{document}